\documentclass[final,5p,authoryear]{elsarticle}

\usepackage{amsmath}
\usepackage{amssymb}
\usepackage{graphicx}
\usepackage{epsfig}
\usepackage[breaklinks]{hyperref}

\journal{Nuclear Physics A}

\def\mnras{MNRAS}
\def\apj{ApJ}

\def\apss{Ap\&SS}

\def\pr{Phys. Rev.}

\def\prc{Phys. Rev. C}
\def\prd{Phys. Rev. D}

\def\nat{Nature}

\begin{document}

\begin{frontmatter}

\title{Uniformly rotating neutron stars in the global and local charge neutrality cases} 

\author[roma,pescara]{Riccardo Belvedere}
\ead{riccardo.belvedere@icra.it}

\author[almaty]{Kuantay Boshkayev}
\ead{kuantay@icra.it}

\author[roma,pescara]{Jorge A. Rueda}
\ead{jorge.rueda@icra.it}

\author[roma,pescara,nizza]{Remo Ruffini}
\ead{ruffini@icra.it}

\address[roma]{Dipartimento di Fisica and ICRA, Sapienza Universita' di Roma\\
P.le Aldo Moro 5, I-00185 Rome, Italy}
\address[pescara]{ICRANet, P.zza della Repubblica 10, I-65122 Pescara, Italy}
\address[nizza]{ICRANet, University of Nice-Sophia Antipolis, 28 Av. de Valrose, 06103 Nice Cedex 2, France}
\address[almaty]{Physical-Technical Faculty, Al-Farabi Kazakh National University\\Al-Farabi ave. 71, 050040 Almaty, Kazakhstan}

\begin{abstract}
In our previous treatment of neutron stars, we have developed the model fulfilling global and not local charge neutrality. In order to implement such a model, we have shown the essential role by the Thomas-Fermi equations, duly generalized to the case of electromagnetic field equations in a general relativistic framework, forming a coupled system of equations that we have denominated Einstein-Maxwell-Thomas-Fermi (EMTF) equations. From the microphysical point of view, the weak interactions are accounted for by requesting the $\beta$ stability of the system, and the strong interactions by using the $\sigma$-$\omega$-$\rho$ nuclear model, where $\sigma$, $\omega$ and $\rho$ are the mediator massive vector mesons. Here we examine the equilibrium configurations of slowly rotating neutron stars by using the Hartle formalism in the case of the EMTF equations indicated above. We integrate these equations of equilibrium for different central densities $\rho_c$ and circular angular velocities $\Omega$ and compute the mass $M$, polar $R_p$ and equatorial $R_{\rm eq}$ radii, angular momentum $J$, eccentricity $\epsilon$, moment of inertia $I$, as well as quadrupole moment $Q$ of the configurations. Both the Keplerian mass-shedding limit and the axisymmetric secular instability are used to construct the new mass-radius relation. We compute the maximum and minimum masses and rotation frequencies of neutron stars. 
We compare and contrast all the results for the global and local charge neutrality cases.
\end{abstract}

\end{frontmatter}

\section{Introduction}\label{sec:1}

We have recently shown \citep{2011PhLB..701..667R,2011NuPhA.872..286R,2012NuPhA.883....1B} that the equations of Tolman-Oppenheimer-Volkoff (TOV) \citep{tolman39,oppenheimer39}, traditionally used to describe the neutron star equilibrium configurations, are superseded once the strong, weak, electromagnetic and gravitational interactions are taken into account. Instead, the Einstein-Maxwell system of equations coupled with the general relativistic Thomas-Fermi equations of equilibrium have to be used; what we called the Einstein-Maxwell-Thomas-Fermi (EMTF) system of equations. While in the TOV approach the condition of local charge neutrality, $n_e(r)=n_p(r)$ is imposed \citep[see e.g.][and references therein]{haenselbook}, the EMTF approach requests the less stringent condition of global charge neutrality, namely 
\begin{equation}
\int \rho_{\rm ch} d^3 r=\int e [n_p(r)- n_e(r)] d^3r = 0,
\end{equation}
where $\rho_{\rm ch}$ is the charge density, $e$ is the fundamental electric charge, and the integral is carried out on the entire volume of the system.

The Lagrangian density taking into account all the interactions include the free-fields terms $\mathcal{L}_g$, $\mathcal{L}_\gamma$, $\mathcal{L}_\sigma$, $\mathcal{L}_\omega$, $\mathcal{L}_\rho$ (respectively for the gravitational, the electromagnetic, and the three mesonic fields), the three fermion species (electrons, protons and neutrons) term $\mathcal{L}_f$ and the interacting part in the minimal coupling assumption, $\mathcal{L}_{\rm int}$ \citep{2011NuPhA.872..286R,2012NuPhA.883....1B}:
\begin{equation}\label{eq:Lagrangian}
\mathcal{L}=\mathcal{L}_{g}+\mathcal{L}_{f}+\mathcal{L}_{\sigma}+\mathcal{L}_{\omega}+\mathcal{L}_{\rho}+\mathcal{L}_{\gamma}+\mathcal{L}_{\rm int} \;,
\end{equation}
where\footnote{We use spacetime metric signature (+,-,-,-) and geometric units $G=c=1$ unless otherwise specified.}
\begin{align*}
\mathcal{L}_g &= -\frac{R}{16 \pi},\quad \mathcal{L}_f = \sum_{i=e, N}\bar{\psi}_{i}\left(i \gamma^\mu D_\mu-m_i \right)\psi_i,\\
\mathcal{L}_{\sigma} &= \frac{\nabla_{\mu}\sigma \nabla^{\mu}\sigma}{2}-U(\sigma),\,
\mathcal{L}_{\omega} = -\frac{\Omega_{\mu\nu}\Omega^{\mu\nu}}{4}+\frac{m_{\omega}^{2} \omega_{\mu} \omega^{\mu}}{2},\\
\mathcal{L}_{\rho} &= -\frac{\mathcal{R}_{\mu\nu}\mathcal{R}^{\mu\nu}}{4}+\frac{m_{\rho}^{2} \rho_{\mu} \rho^{\mu}}{2},\quad\mathcal{L}_{\gamma} = -\frac{F_{\mu\nu}F^{\mu\nu}}{16\pi},\\
\mathcal{L}_{\rm int} &= -g_{\sigma} \sigma \bar{\psi}_N \psi_N - g_{\omega} \omega_{\mu} J_{\omega}^{\mu}-g_{\rho}\rho_{\mu}J_{\rho}^{\mu} + e A_{\mu} J_{\gamma,e}^{\mu} \\
&-e A_{\mu} J_{\gamma,N}^{\mu},
\end{align*}
where the description of the strong interactions between the nucleons is made through the $\sigma$-$\omega$-$\rho$ nuclear model in the version of \cite{boguta77}. Thus $\Omega_{\mu\nu}\equiv\partial_{\mu}\omega_{\nu}-\partial_{\nu}\omega_{\mu}$, $\mathcal{R}_{\mu\nu}\equiv\partial_{\mu}\rho_{\nu}-\partial_{\nu}\rho_{\mu}$,
$F_{\mu\nu}\equiv\partial_{\mu}A_{\nu}-\partial_{\nu}A_{\mu}$ are the field strength tensors for the
$\omega^{\mu}$, $\rho$ and $A^{\mu}$ fields respectively, $\nabla_\mu$ stands for covariant derivative and $R$ is the Ricci scalar. We adopt the Lorenz gauge for the fields $A_\mu$, $\omega_\mu$, and $\rho_\mu$. The self-interaction scalar field potential is $U(\sigma)$, $\psi_N$ is the nucleon isospin doublet, $\psi_e$ is the electronic singlet, $m_i$ states for the mass of each particle-specie and $D_\mu = \partial_\mu + \Gamma_\mu$, being $\Gamma_\mu$ the Dirac spin connections. The conserved currents are $J^{\mu}_{\omega} = \bar{\psi}_N \gamma^{\mu}\psi_N$, $J^{\mu}_{\rho} = \bar{\psi}_N \tau_3\gamma^{\mu}\psi_N$, $J^{\mu}_{\gamma, e} = \bar{\psi}_e \gamma^{\mu}\psi_e$, and $J^{\mu}_{\gamma, N} = \bar{\psi}_N(1/2)(1+\tau_3)\gamma^{\mu}\psi_N$, being $\tau_3$ the particle isospin.

The nuclear model is fixed once the values of the coupling constants and the masses of the three mesons are fixed: for instance in the NL3 parameter set \cite{lalazissis97} used in \citep{2012NuPhA.883....1B} and in this work we have $m_\sigma=508.194$ MeV, $m_\omega=782.501$ MeV, $m_\rho=763.000$ MeV, $g_\sigma=10.2170$, $g_\omega=12.8680$, $g_\rho=4.4740$, plus two constants that give the strength of the self-scalar interactions, $g_2=-10.4310$ fm$^{-1}$ and $g_3=-28.8850$.

\begin{figure}[!hbtp]
\centering
\includegraphics[width=0.8\hsize,clip]{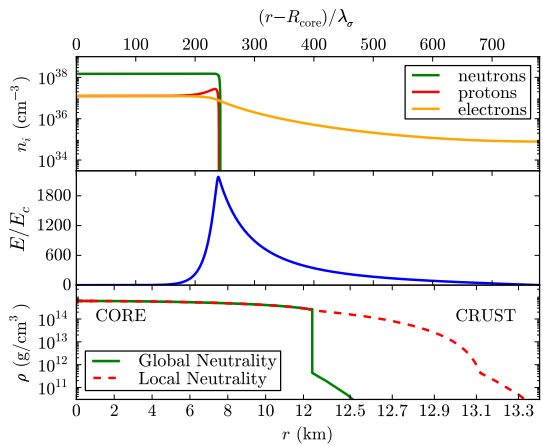}
\caption{In the top and center panels we show the neutron, proton, electron densities and the electric field in units of the critical electric field $E_c$ in the core-crust transition layer, whereas in the bottom panel we show a specific example of a density profile inside a neutron star. In this plot we have used for the globally neutral case a density at the edge of the crust equal to the neutron drip density, $\rho_{\rm drip}\sim 4.3\times 10^{11}$ g cm$^{-3}$.}\label{fig:Model}
\end{figure}

From the equations of motion of the above Lagrangian we obtain the EMTF equations \citep[see][for details]{2011NuPhA.872..286R,2012NuPhA.883....1B}. The solution of the EMTF coupled differential equations leads to a new structure of the star, as shown in Fig~\ref{fig:Model}: a positively charged core at supranuclear densities, $\rho>\rho_{\rm nuc}\sim 2.7\times 10^{14}$ g cm$^{-3}$, surrounded by an electron distribution of thickness $\gtrsim \hbar/(m_e c)$ and, at lower densities $\rho<\rho_{\rm nuc}$, a neutral ordinary crust. 

The thermodynamic equilibrium is ensured by the constancy of the particle Klein potentials \cite{klein49} generalized to the presence of electrostatic and strong fields \citep{2011PhLB..701..667R,2011NuPhA.872..286R,2012NuPhA.883....1B}
\begin{equation}\label{eq:klein}
\frac{1}{u^t}\,[\mu_i + (q_i A_\alpha + g_\omega \omega_\alpha  + g_\rho \tau_{3,i} \rho_\alpha) u^\alpha]={\rm constant},
\end{equation}
where the subscript $i$ stands for each kind of particle, $\mu_i$ is the particle chemical potential, and $q_i$ is the particle electric charge. In the static case only the time components of the vector fields, $A_0$, $\omega_0$, $\rho_0$ are present. In the above equation $u^t=(g_{tt})^{-1/2}$ is the time component of the fluid four-velocity which satisfies $u_\alpha u^\alpha =1$; $g_{tt}$ is the t--t component of the spherically symmetric metric
\begin{equation}\label{eq:metric1}
ds^2=e^{\nu} dt^2-e^{\lambda} dr^2-dr^2-r^2 (d\theta^2+\sin^2\theta d\phi^2)\;.
\end{equation}

The equilibrium conditions (\ref{eq:klein}) lead to a discontinuity in the density at the core-crust transition and, correspondingly, an overcritical electric field $\sim (m_\pi/m_e)^2 E_c$, where $E_c=m^2_e c^3/(e \hbar)\sim 1.3\times 10^{16}$ Volt cm$^{-1}$, appears in the core-crust boundary interface. The constancy of the Klein potentials is necessary to fulfill the requirement of thermodynamical equilibrium, together with the constancy of the gravitationally red-shifted temperature (Tolman condition) \citep{1930PhRv...35..904T,klein49}, if finite temperatures are considered \citep[see e.g.][]{2011NuPhA.872..286R}. In particular, the continuity of the electron Klein potential leads to a decreasing of the electron chemical potential $\mu_e$ and density at the core-crust boundary interface. They reach values $\mu^{\rm crust}_e < \mu^{\rm core}_e$ and $\rho_{\rm crust}<\rho_{\rm core}$ at the base of the crust, where global charge neutrality is achieved.

\begin{figure}[!hbtp]
\centering
\includegraphics[width=0.8\hsize,clip]{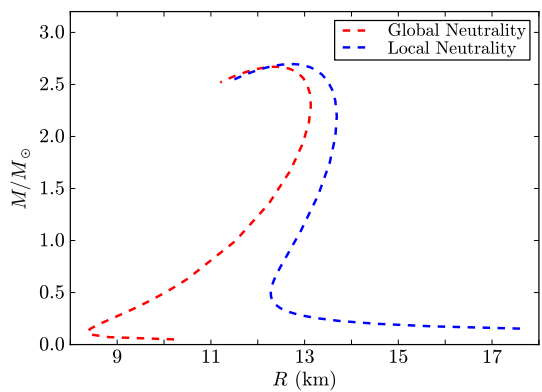}
\caption{Neutron star mass-radius relation in the static (non-rotating) case for both global and local charge neutrality configurations \citep[see][for details]{2012NuPhA.883....1B}. In this plot we have used for the globally neutral case a density at the edge of the crust equal to the neutron drip density, $\rho_{\rm drip}\sim 4.3\times 10^{11}$ g cm$^{-3}$.}\label{fig:MRstat}
\end{figure}

As we showed in \citep{2012NuPhA.883....1B}, the solution of this new set of equilibrium equations leads to a more compact neutron star with a less massive and thiner crust. Consequently, it leads to a new mass-radius relation which markedly differs from the one given by the solution of the TOV equations in the case of local charge neutrality; see Fig.~\ref{fig:MRstat}.

We extend in this work the previous results to the case when the neutron star is rotating as a rigid body. To this end we use the Hartle approach \citep{1967ApJ...150.1005H} which solves the Einstein equations accurately up to second order approximation in the angular velocity of the star, $\Omega$ (see next section \ref{sec:2} for details).

In this rotating case, the condition of the constancy of the particle Klein potential has the same form as Eq.~(\ref{eq:klein}), but the fluid inside the star now moves with a four-velocity of a rigid rotating body, $u^\alpha=(u^t,0,0,u^\phi)$, with (see \cite{HS1967} and \ref{app:1}, for details)
\begin{equation}
u^t=(g_{tt}+2\Omega\,g_{t\phi}+\Omega^2\,g_{\phi \phi})^{-1/2},\qquad u^\phi=\Omega u^t,
\end{equation}
where $\phi$ is the azimuthal angular coordinate with respect to which the metric is symmetric, namely the metric is independent of $\phi$ (axial symmetry). The metric functions $g_{\alpha \beta}$ are now given by Eq.~(\ref{eq:rotmetric}) below. It is then clear that in a frame comoving with the rotating star, $u^t=(g_{tt})^{-1/2}$, and the Klein equilibrium condition becomes the same as Eq.~(\ref{eq:klein}), as expected. 

We applied the Hartle formalism to the seed static solution obtained from the integration of the EMTF equations \citep{2012NuPhA.883....1B}. For the construction of the new mass-radius relation we take into account the Keplerian mass-shedding limit and the secular axisymmetric instability (see section \ref{sec:3}). We compute in section~\ref{sec:4} the mass $M$, polar $R_p$ and equatorial $R_{\rm eq}$ radii, and angular momentum $J$, as a function of the central density and the rotation angular velocity $\Omega$ of stable neutron stars both in the globally and locally neutral cases. Based on the criteria of equilibrium we calculate the maximum stable neutron star mass and from the gravitational binding energy of the configurations establish the minimum mass under which the neutron star becomes gravitationally unbound. We construct in section~\ref{sec:5} the new neutron star mass-radius relation. In section~\ref{sec:6} we calculate the moment of inertia as a function of the central density and total mass of the neutron star. The eccentricity $\epsilon$, the rotational to gravitational energy ratio $T/W$, and quadrupole moment $Q$ are shown in section~\ref{sec:7}. The observational constraints on the mass-radius relation are discussed in section \ref{sec:8}. We finally summarize the results in section \ref{sec:9}.

\section{Hartle slow rotation approximation}\label{sec:2}

In his pioneering work, \cite{1967ApJ...150.1005H} computed the equilibrium equations of slowly rotating stars in the context of General Relativity. The solutions of the Einstein equations are obtained through a perturbative method, expanding the metric functions up to the second order in the angular velocity $\Omega$. Under this assumption the structure of compact objects can be approximately described by the total mass $M$, angular momentum $J$ and quadrupole moment $Q$. The slow rotation regime implies that the perturbations owing to the rotation are relatively small with respect to the known non-rotating geometry. The interior solution is derived by solving numerically a system of ordinary differential equations for the perturbation functions. The exterior solution for the vacuum surrounding the star, can be written analytically in terms of $M$, $J$, and $Q$ \citep[see][for details]{1967ApJ...150.1005H,1968ApJ...153..807H}. The numerical values for all the physical quantities are derived by matching the interior and the exterior solution on the border of the star.

The spacetime metric for the rotating configuration up to the second order of $\Omega$ is given by \citep{1967ApJ...150.1005H}
\begin{eqnarray}\label{eq:rotmetric}
ds^2 &=& e^{\nu}\left(1+2h\right)dt^2-e^{\lambda}\left[1+\frac{2m}{r-2 M_0}\right]dr^2  \nonumber\\
&-& r^2\left(1+2k\right)\left[d\theta^2+\sin^2\theta\left(d\phi-\omega dt\right)^2\right]\, ,
\end{eqnarray}
where $\nu=\nu(r)$, $\lambda=\lambda(r)$, and $M_0=M^{J=0}(r)$ are the metric functions and mass profiles of the corresponding seed non-rotating star with the same central density as the rotating one; see Eq.~(\ref{eq:metric1}). The functions $h=h(r,\theta)$, $m=m(r,\theta)$, $k=k(r,\theta)$ and the fluid angular velocity in the local inertial frame, $\omega=\omega(r)$, have to be calculated from the Einstein equations. Expanding up to the second order the metric in spherical harmonics we have
\begin{eqnarray}\label{eq:HarmonicExp}
&&h(r,\theta)=h_0(r)+h_2(r)P_2(\cos\theta) \;,\\
&&m(r,\theta)=m_0(r)+m_2(r)P_2(\cos\theta) \;,\\
&&k(r,\theta)=k_0(r)+k_2(r)P_2(\cos\theta) \;,
\end{eqnarray}
where $P_2(cos\theta)$ is the Legendre polynomial of second order. Because the metric does not change under transformations of the type $r\rightarrow f(r)$, we can assume $k_0(r)=0$. 

The functions $h=h(r,\theta)$, $m=m(r,\theta)$, $k=k(r,\theta)$ have analytic form in the exterior (vacuum) spacetime and they can be found in \ref{app:1}. The mass, angular momentum, and quadrupole moment are computed from the matching condition between the interior and exterior metrics.

First the angular momentum is computed. It is introduced the angular velocity of the fluid relative to the local inertial frame, $\bar{\omega}(r)=\Omega-\omega(r)$. It can be shown from the Einstein equations at first order in $\Omega$ that $\bar{\omega}$ satisfies the differential equation
\begin{equation}\label{eq:baromega}
\frac{1}{r^4}\frac{d}{dr}\left( r^4 j \frac{d\bar{\omega}}{dr} \right)+\frac{4}{r}\frac{d j}{dr}\bar{\omega}=0\;,
\end{equation}
where $j(r)=e^{-(\nu+\lambda)/2}$ with $\nu$ and $\lambda$ the metric functions of the seed non-rotating solution (\ref{eq:metric1}). 

From the matching equations, the angular momentum of the star results to be given by
\begin{equation}\label{eq:J}
J = \frac{1}{6}R^4\left(\frac{d\bar{\omega}}{dr}\right)_{r=R}\;,
\end{equation}
so the angular velocity $\Omega$ is related to the angular momentum as
\begin{equation}\label{eq:Jomega}
\Omega = \bar{\omega}(R)+\frac{2 J}{R^3}\;.
\end{equation}

The total mass of the rotating star, $M$, is given by 
\begin{equation}\label{eq:Mrot}
M = M_0+\delta M\;,\qquad \delta M = m_0(R)+J^2/R^3\,,
\end{equation}
where $\delta M$ is the contribution to the mass owing to rotation. The second order functions $m_0$ and $p_0^*$ (related to the pressure perturbation) are computed from the solution of the differential equation
\begin{align}
\frac{d m_0}{dr}&=4\pi r^2 \frac{d{\cal E}}{dP} ({\cal E}+P) p_0^* + \frac{1}{12}j^2 r^4 \left(\frac{d\bar{\omega}}{dr}\right)^2 \nonumber\\
&-\frac{1}{3}\frac{dj^2}{dr}r^3 \bar{\omega}^2\;,\\
\frac{d p_0^*}{dr}&=-\frac{m_0 (1+8 \pi r^2 P)}{(r-2 M_0)^2}-\frac{4\pi r^2 ({\cal E}+P)}{(r-2 M_0)}p_0^* \nonumber \\
&+ \frac{1}{12}\frac{j^2 r^4}{(r-2 M_0)}\left(\frac{d\bar{\omega}}{dr}\right)^2 + \frac{1}{3}\frac{d}{dr} \left(\frac{r^3j^2\bar{\omega}^2}{r-2 M_0}\right)\;,
\end{align}
where ${\cal E}$ and $P$ are the total energy-density and pressure.

Turning to the quadrupole moment of the neutron star, it is given by 
\begin{equation}\label{eq:Q}
Q=\frac{J^2}{M_0}+\frac{8}{5}{\cal K} M_0^3\;,
\end{equation}
where ${\cal K}$ is a constant of integration. This constant is fixed from the matching of the second order function $h_2$ obtained in the interior from
\begin{align}
\frac{d k_2}{dr}&=-\frac{d h_2}{dr}-h_2\frac{d\nu}{dr}+\left(\frac{1}{r}+\frac{1}{2}\frac{d\nu}{dr}\right)\bigg[-\frac{1}{3}r^3\bar{\omega}^2\frac{dj^2}{dr} \nonumber\\
&+ \frac{1}{6}r^4 j^2 \left(\frac{d\bar{\omega}}{dr}\right)^2\bigg]\;,\\
\frac{d h_2}{dr}&=h_2\bigg\{-\frac{d\nu}{dr}+\frac{r}{r-2 M_0}\left(\frac{d\nu}{dr}\right)^{-1}\bigg[8\pi({\cal E}+P)\nonumber\\
&-\frac{4 M_0}{r^3}\bigg] \bigg\}-\frac{4 (k_2+h_2)}{r (r-2 M_0)}\left(\frac{d\nu}{dr}\right)^{-1}\nonumber\\
&+\frac{1}{6}\bigg[\frac{r}{2}\frac{d\nu}{dr}-\frac{1}{r-2 M_0}\left(\frac{d\nu}{dr}\right)^{-1}\bigg]r^3j^2\left(\frac{d\bar{\omega}}{dr}\right)^2\nonumber\\
&-\frac{1}{3}\bigg[\frac{r}{2}\frac{d\nu}{dr}+\frac{1}{r-2 M_0}\left(\frac{d\nu}{dr}\right)^{-1}\bigg]r^2 \bar{\omega}^2\frac{dj^2}{dr}\;,
\end{align}
with its exterior counterpart (see \cite{1967ApJ...150.1005H} and \ref{app:1}).

It is worth to underline that the influence of the induced magnetic field owing to the rotation of the charged core of the neutron star in the globally neutral case is negligible \citep{2012IJMPS..12...58B}. In fact, for a rotating neutron star of period $P=10$ ms and radius $R\sim10$ km, the radial component of the magnetic field $B_r$ in the core interior reaches its maximum at the poles with a value $B_r\sim 2.9\times10^{-16}B_c$, where $B_c=m_e^2c^3/(e\hbar)\approx 4.4\times10^{13}$ G is the critical magnetic field for vacuum polarization. The angular component of the magnetic field $B_\theta$, instead, has its maximum value at the equator and, as for the radial component, it is very low in the interior of the neutron star core, i.e. $|B_\theta|\sim 2.9\times 10^{-16}B_c$. In the case of a sharp core-crust transition as the one studied by \cite{2012NuPhA.883....1B} and shown in Fig.~\ref{fig:Model}, this component will grow in the transition layer to values of the order of $|B_\theta|\sim 10^2 B_c$ \citep[see][for further details]{2012IJMPS..12...58B}. However, since we are here interested in the macroscopic properties of the neutron star, we can ignore at first approximation the presence of electromagnetic fields in the macroscopic regions where they are indeed very small, and safely apply the original Hartle formulation without any generalization.

\section{Stability of uniformly rotating neutron stars}\label{sec:3}

\subsection{Secular axisymmetric instability}\label{subsec:3.1}

In a sequence of increasing central density in the $M$-$\rho_c$ curve, $\rho_c\equiv \rho(0)$, the maximum mass of a non-rotating neutron star is defined as the first maximum of such a curve, namely the point where $\partial M$/$\partial \rho_c=0$. This derivative defines the secular instability point, and, if the perturbation obeys the same equation of state (EOS) as the equilibrium configuration, it coincides also with the dynamical instability point \citep[see e.g.][]{shapirobook}. In the rotating case, the situation becomes more complicated and in order to find the axisymmetric dynamical instability points, the perturbed solutions with zero frequency modes (the so-called neutral frequency line) have to be calculated. \cite{1988ApJ...325..722F} however, following the works of \cite{1981ApJ...249..254S,1982ApJ...257..847S}, described a turning-point method to obtain the points at which secular instability is reached by uniformly rotating stars. In a constant angular momentum sequence, the turning point is located in the maximum of the mass-central density relation, namely the onset of secular axisymmetric instability is given by
\begin{equation}\label{eq:TurningPoint}
\left[\frac{\partial M\left(\rho_c,J\right)}{\partial\rho_c}\right]_{J=\rm constant}=0 \;,
\end{equation}
and once the secular instability sets in, the star evolves quasi-stationarily until it reaches a point of dynamical instability where gravitational collapse sets in \citep{2003LRR.....6....3S}.

The above equation defines an upper limit for the mass at a given $J$ for a uniformly rotating star, however this criterion is a sufficient but not necessary condition for the instability. This means that all the configurations with the given angular momentum $J$ on the right side of the turning point defined by Eq.~(\ref{eq:TurningPoint}) are secularly unstable, but it does not imply that the configurations on the left side of it are stable. An example of dynamically unstable configurations on the left side of the turning-point limiting boundary in neutron stars was recently shown in \citep{2011MNRAS.416L...1T}, for a specific EOS.

\subsection{Keplerian mass-shedding instability}\label{subsec:3.2}

The maximum velocity for a particle to remain in equilibrium on the equator of a star, kept bound by the balance between gravitational and centrifugal force, is the Keplerian velocity of a free particle computed at the same location. As shown, for instance in \citep{2003LRR.....6....3S}, a star rotating at Keplerian rate becomes unstable due to the loss of mass from its surface. The mass shedding limiting angular velocity of a rotating star is the Keplerian angular velocity evaluated at the equator, $r=R_{\rm eq}$, i.e. $\Omega_K^{J\neq0}=\Omega_K(r=R_{\rm eq})$. \cite{Friedman1986} introduced a method to obtain the maximum possible angular velocity of the star before reaching the mass-shedding limit; however \cite{2008AcA....58....1T} and \cite{BBRS2013}, showed a simpler way to compute the Keplerian angular velocity of a rotating star. They showed that the mass-shedding angular velocity, $\Omega_K^{J\neq0}$, can be computed as the orbital angular velocity of a test particle in the external field of the star and corotating with it on its equatorial plane at the distance $r=R_{\rm eq}$. For the Hartle external solution, this is given by
\begin{equation}\label{eq:omegaKep}
\Omega_{K}^{J\neq0}(r)=\sqrt{\frac{M}{r^3}}\left[1- j F_{1}(r)+j^2F_{2}(r)+q F_{3}(r)\right] \;,
\end{equation}
where $j=J/M^2$ and $q=Q/M^3$ are the dimensionless angular momentum and quadrupole moment. Further details and the analytical expression of the functions $F_i$ can be found in \ref{app:1}.

\subsection{Gravitational binding energy}\label{subsec:3.3}

Besides the above stability requirements, one should check if the neutron star is gravitationally bound. In the non-rotating case, the binding energy of the star can be computed as
\begin{equation}\label{eq:W0}
W_{J=0}=M_0-M^0_{\rm rest}\;,\qquad M^0_{\rm rest}=m_b A_{J=0}\;,
\end{equation}
where $M^0_{\rm rest}$ is the rest-mass of the star, $m_b$ is the rest-mass per baryon, and $A_{J=0}$ is the total number of baryons inside the star. So the non-rotating star is considered bound if $W_{J=0}<0$.

In the slow rotation approximation the total binding energy is given by \citep{1968ApJ...153..807H}
\begin{equation}\label{eq:Wrot}
W_{J\neq 0}=W_{J=0}+\delta W\;,\qquad  \delta W=\frac{J^2}{R^3}-\int_{0}^R 4\pi r^2 B(r)dr\;,
\end{equation}
where 
\begin{align}
&B(r)=({\cal E}+P)p^*_0\bigg\{ \frac{d{\cal E}}{dP}\left[\left(1-\frac{2 M}{r}\right)^{-1/2}-1\right]\nonumber\\
&- \frac{d u}{dP}\left(1-\frac{2 M}{r}\right)^{-1/2}\bigg\}+({\cal E}-u)\left(1-\frac{2 M}{r}\right)^{-3/2}\bigg[\frac{m_0}{r}\nonumber \\
&+\frac{1}{3}j^2r^2\bar{\omega}^2\bigg]-\frac{1}{4\pi r^2}\left[ \frac{1}{12}j^2 r^4 \left(\frac{d\bar{\omega}}{dr}\right)^2-\frac{1}{3}\frac{dj^2}{dr}r^3 \bar{\omega}^2 \right]\;,
\end{align}
where $u={\cal E}-m_b n_b$ is the internal energy of the star, with $n_b$ the baryon number density.

We will therefore request that the binding energy be negative, namely $W_{J\neq 0}<0$. As we will show below in section~\ref{sec:4.2.2}, this condition leads to a minimum mass for the neutron star under which the star becomes gravitationally unbound.

\section{Structure of uniformly rotating neutron stars}\label{sec:4}

We show now the results of the integration of the Hartle equations for the globally and locally charge neutrality neutron stars; see e.g.~Fig.~\ref{fig:Model}. Following \cite{2012NuPhA.883....1B}, we adopt, as an example, globally neutral neutron stars with a density at the edge of the crust equal to the neutron drip density, $\rho_{\rm crust}=\rho_{\rm drip}\approx 4.3\times 10^{11}$ g cm$^{-3}$.

\subsection{Secular instability boundary}\label{sec:4.1}

\begin{figure}[!hbtp]
\centering
\includegraphics[width=0.8\hsize,clip]{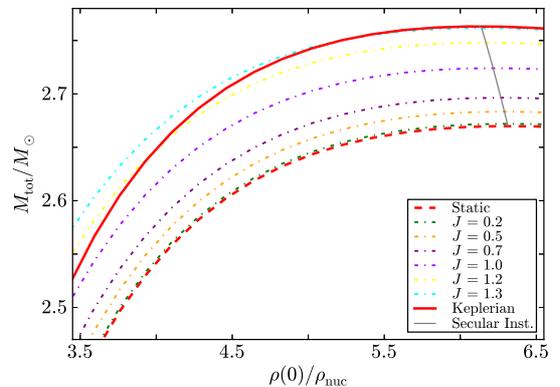}
\caption{Total mass versus central density of globally neutral neutron stars. The solid line represents the configuration with Keplerian angular velocity, the dashed line represents the static configuration, the dotted-dashed lines represent the $J$-constant sequences (in units of 10$^{11}$ cm$^2$). The gray line joins all the turning points of the $J$-constant sequences, so it defines the secular instability boundary.}\label{fig:MtotvsRhoG}
\end{figure}

In Fig.~\ref{fig:MtotvsRhoG} we show the mass-central density curve for globally neutral neutron stars in the region close to the axisymmetric stability boundaries. Specifically we show some $J$-constant sequences to show that indeed along each of these curves there exist a maximum mass point (turning point). The line joining all the turning points defines the secular instability limit. In Fig.~\ref{fig:MtotvsRhoG} the axisymmetric stable zone is on the left side of the instability line.

\begin{figure}[!hbtp]
\centering
\includegraphics[width=0.8\hsize,clip]{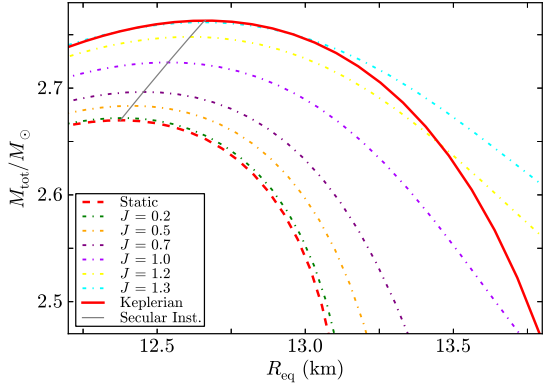}
\caption{Total mass versus equatorial radius of globally neutral neutron stars. The solid line represents the configuration with Keplerian angular velocity, the dashed line represents the static configuration, the dotted-dashed lines represent the $J$-constant sequences (in units of 10$^{11}$ cm$^2$). The gray curve joins all the turning points of the $J$-constant sequences, so it defines the secular instability boundary.}\label{fig:MtotvsReqG}
\end{figure}

Clearly we can transform the mass-central density relation in a mass-radius relation. In Fig.~\ref{fig:MtotvsReqG} we show the mass versus the equatorial radius of the neutron star that correspond to the range of densities of Fig.~\ref{fig:MtotvsRhoG}. In this plot the stable zone is on the right side of the instability line. 

We can construct a fitting curve joining the turning points of the $J$-constant sequences line which determines the secular axisymmetric instability boundary. Defining $M_{{\rm max},0}$ as the maximum stable mass of the non-rotating neutron star constructed with the same EOS, we find that for globally neutral configurations the instability line is well fitted by the function
\begin{align}\label{eq:SecularG}
\frac{M^{\rm GCN}_{\rm sec}}{M_\odot}&=21.22-6.68\frac{M_{{\rm max},0}^{\rm GCN}}{M_\odot}\nonumber \\
&-\left(77.42-28\frac{M_{{\rm max},0}^{\rm GCN}}{M_\odot}\right)\left(\frac{R_{\rm eq}}{10\,{\rm km}}\right)^{-6.08}\;,
\end{align}
where $12.38\,{\rm km}\lesssim R_{\rm eq}\lesssim 12.66\,{\rm km}$, and $M_{{\rm max},0}^{\rm GCN}\approx 2.67 M_\odot$.

\begin{figure}[!hbtp]
\centering
\includegraphics[width=0.8\hsize,clip]{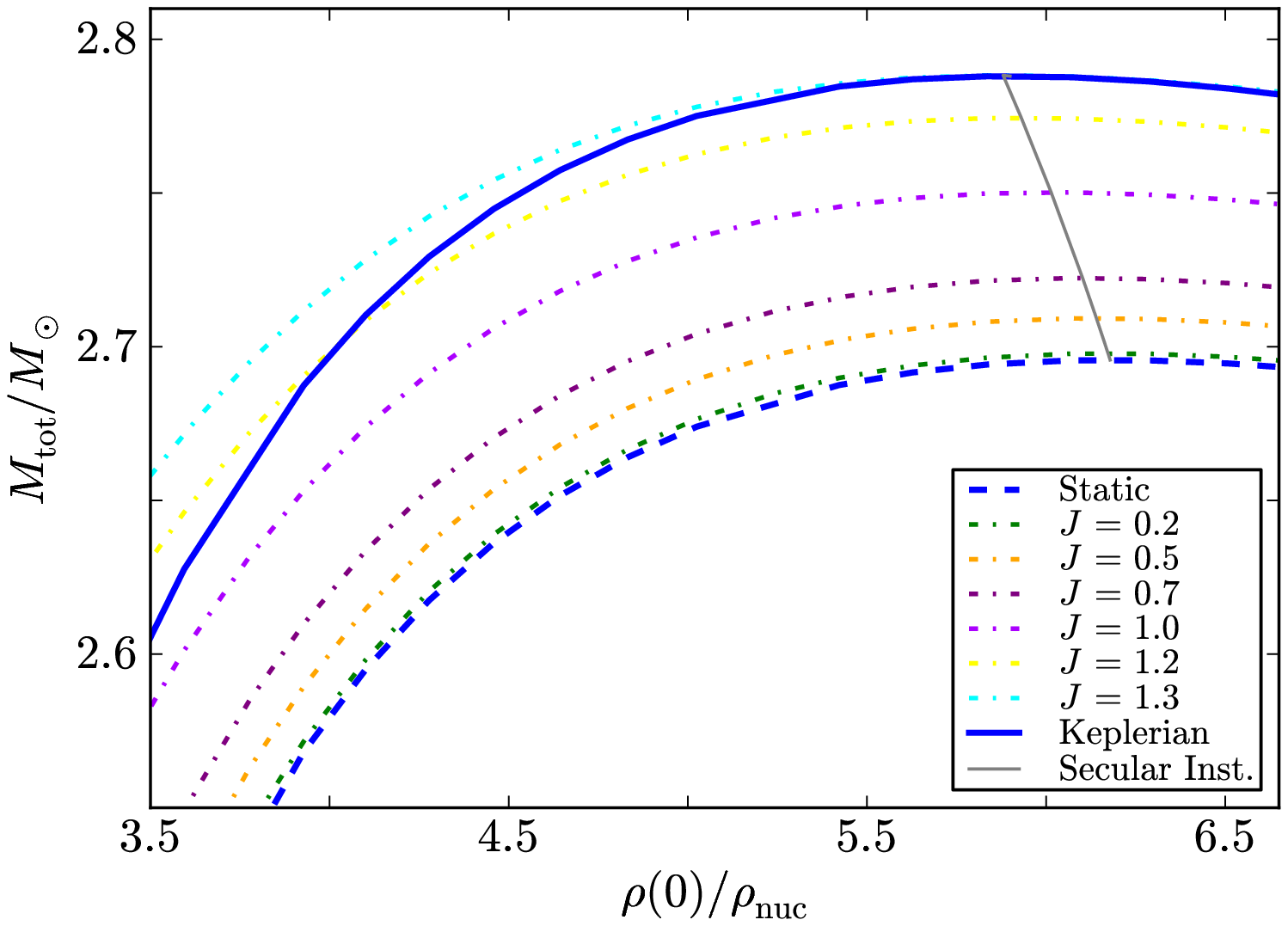}
\caption{Total mass versus central density of locally neutral neutron stars. The solid line represents the configuration with Keplerian angular velocity, the dashed line represents the static configuration, the dotted-dashed lines represent the $J$-constant sequences (in units of 10$^{11}$ cm$^2$). The gray line joins all the turning points of the $J$-constant sequences, so it defines the secular instability boundary.}\label{fig:MtotvsRhoL}
\end{figure}

\begin{figure}[!hbtp]
\centering
\includegraphics[width=0.8\hsize,clip]{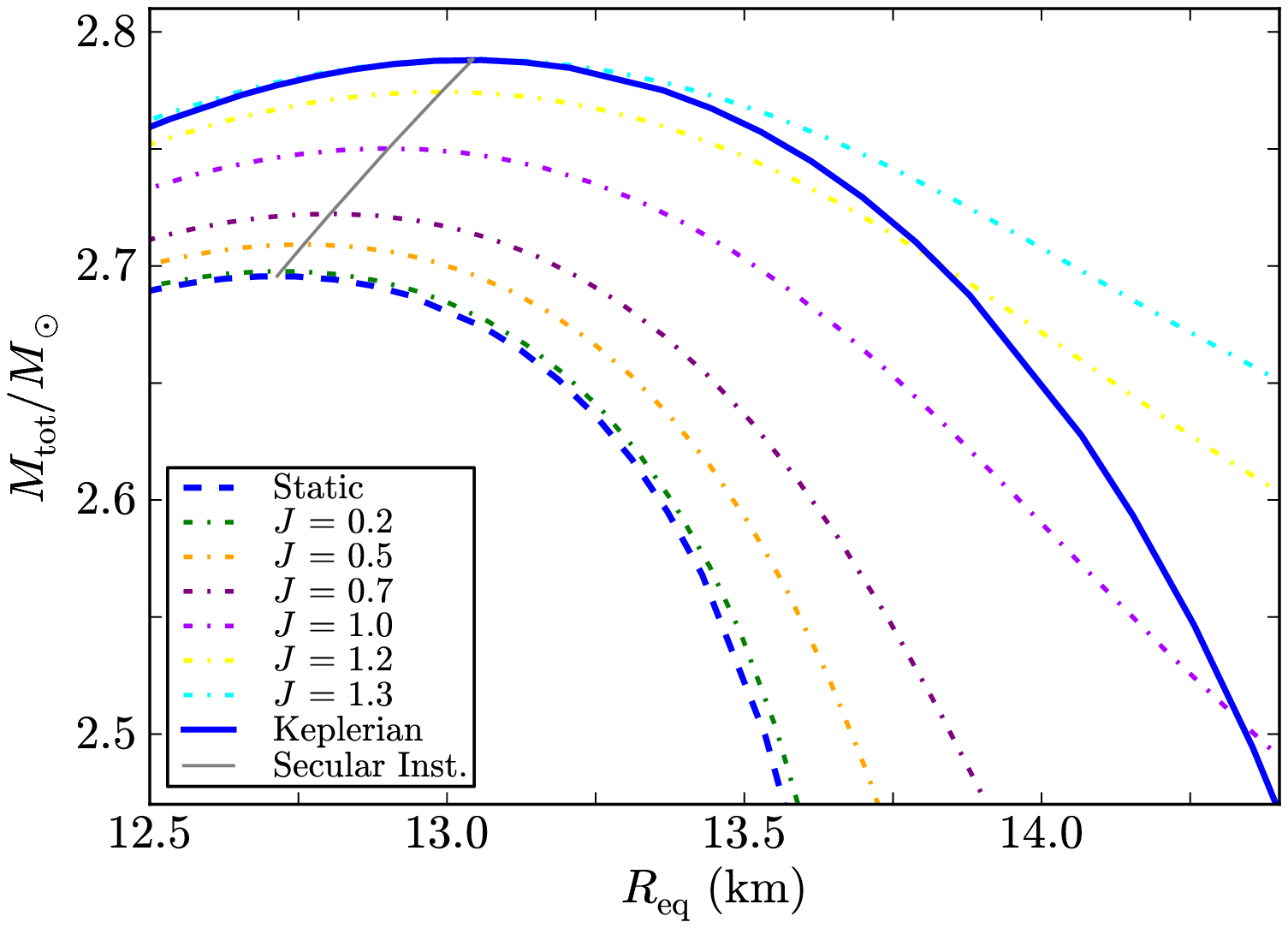}
\caption{Total mass versus equatorial radius of locally neutral neutron stars. The solid line represents the configuration with Keplerian angular velocity, the dashed line represents the static configuration, the dotted-dashed lines represent the $J$-constant sequences (in units of 10$^{11}$ cm$^2$). The gray curve joins all the turning points of the $J$-constant sequences, so it defines the secular instability boundary.}\label{fig:MtotvsReqL}
\end{figure}

The turning points of locally neutral configurations in the mass-central density plane are shown in Fig.~\ref{fig:MtotvsRhoL}. the corresponding mass-equatorial radius plane is plotted in Fig.~\ref{fig:MtotvsReqL}.

For locally neutral neutron stars, the secular instability line is fitted by

\begin{align}\label{eq:SecularL}
\frac{M^{\rm LCN}_{\rm sec}}{M_\odot}&=20.51-6.35\frac{M_{{\rm max},0}^{\rm LCN}}{M_\odot}\nonumber \\
&-\left(80.98-29.02\frac{M_{{\rm max},0}^{\rm LCN}}{M_\odot}\right)\left(\frac{R_{\rm eq}}{10\,{\rm km}}\right)^{-5.71}\;,
\end{align}
where $12.71\,{\rm km}\lesssim R_{\rm eq}\lesssim 13.06\,{\rm km}$, and $M_{\rm max,0}^{\rm LCN}\approx 2.70 M_\odot$. 

\subsection{Keplerian mass-shedding sequence}\label{sec:4.2}

We turn now to analyze in detail the behavior of the different properties of the neutron star along the Keplerian mass-shedding sequence. For the sake of reference we have indicated in the following plots stars with the selected masses $M\approx[1,1.4,2.04,2.5]\,M_\odot$. The cyan star indicates the fastest observed pulsar, PSR J1748--2446ad \cite{hessels06}, with a rotation frequency of $f\approx 716$ Hz. The gray filled circles indicate the last stable configuration of the Keplerian sequence, namely the point where the Keplerian and the secular stability boundaries cross each other.

\subsubsection{Maximum mass and rotation frequency}\label{sec:4.2.1}

\begin{figure}[!hbtp]
\centering
\includegraphics[width=0.8\hsize]{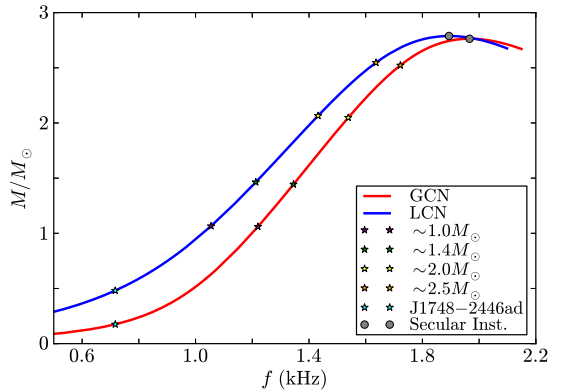}
\caption{Total mass versus rotational Keplerian frequency both for the global (red) and local (blue) charge neutrality cases.}\label{fig:Mtotvsf}
\end{figure}

The total mass of the rotating star is computed from Eq.~(\ref{eq:Mrot}). In Fig.~\ref{fig:Mtotvsf} is shown the total mass of the neutron star as a function of the rotation frequency for the Keplerian sequence. It is clear that for a given mass, the rotational frequency is higher for a globally neutral neutron star with respect to the locally neutral one. 

The configuration of maximum mass, $M^{J\neq 0}_{\rm max}$, occurs along the Keplerian sequence, and it is found before the secular instability line crosses the Keplerian curve. Thus, the maximum mass configuration is secularly stable. This implies that the configuration with maximum rotation frequency, $f_{\rm max}$, is located beyond the maximum mass point, specifically at the crossing point between the secular instability and the Keplerian mass-shedding sequence. The results are summarized in Table~\ref{tab:MRfmax}.

It is important to discuss briefly the validity of the present perturbative solution for the computation of the properties of maximally rotating neutron stars. The expansion of the radial coordinate of a rotating configuration $r(R,\theta)$ in powers of angular velocity is written as \citep{1967ApJ...150.1005H}
\begin{equation}
r=R+\xi(R,\theta)+O(\Omega^4) \; ,
\end{equation}
where $\xi$ is the difference in the radial coordinate, $r$, between a point located at the polar angle $\theta$ on the surface of constant density $\rho(R)$ in the rotating configuration, and the point located at the same polar angle on the same constant density surface in the non-rotating configuration.
In the slow rotation regime, the fractional displacement of the surfaces of constant density due to the rotation have to be small, namely $\xi(R,\theta)/R\ll 1$, where $\xi(R,\theta)=\xi_0(R)+\xi_2(R)P_2(\cos\theta)$ and $\xi_0(R)$ and $\xi_2(R)$ are function of $R$, proportional to $\Omega^2$. From Table~\ref{tab:MRfmax}, we can see that the configuration with the maximum possible rotation frequency has a maximum fractional displacement $\delta R_{\rm eq}^{\rm max}=\xi(R,\pi/2)/R$ as low as $\approx 2\%$ and $\approx 3\%$, for the globally and locally neutral neutron stars respectively.

In this line, it is worth to quote the results of \cite{benhar05}, who showed that the inclusion of a third-order expansion $\Omega^3$ in the Hartle's method improves the value of the maximum rotation frequency by less than 1\% for different EOS. The reason for this is that as mentioned above, along the Keplerian sequence the deviations from sphericity decrease with density and frequency (see Figs.~\ref{fig:eccvsf} and \ref{fig:eccvsrho}), which ensures the accuracy of the perturbative solution.

Turning to the increase of the maximum mass, \cite{1992ApJ...390..541W} showed that the mass of maximally rotating neutron stars, computed with the Hartle's second order approximation, is accurate within an error as low as $\lesssim 4\%$.

\begin{table}
\centering
\begin{tabular}{c c c}
& Global Neutrality & Local Neutrality \\
\hline
$M^{J=0}_{\rm max}$ $(M_\odot)$ & 2.67 & 2.70\\
$R^{J=0}_{\rm max}$ (km) & 12.38 & 12.71\\
$M^{J\neq 0}_{\rm max}$ $(M_\odot)$ & 2.76 & 2.79\\
$R^{J\neq 0}_{\rm max}$ (km) & 12.66 & 13.06\\
$\delta M_{\rm max}$ & 3.37\% & 3.33\%\\
$\delta R_{\rm eq}^{\rm max}$ & 2.26\% & 2.75\%\\
$f_{\rm max}$ (kHz)& 1.97 & 1.89\\
$P_{\rm min}$ (ms)& 0.51 & 0.53\\
\hline
\end{tabular}
\caption{$M^{J=0}_{\rm max}$ and $R^{J=0}_{\rm max}$: maximum mass and corresponding radius of non-rotating stars as computed in \citep{2012NuPhA.883....1B}; $M^{J=0}_{\rm max}$ and $R^{J=0}_{\rm max}$: maximum mass and corresponding radius of rotating stars; $\delta M_{\rm max}$ and $\delta R_{\rm eq}^{\rm max}$: increase in mass and radius of the maximum mass configuration with respect to its non-rotating counterpart; $f_{\rm max}$ and $P_{\rm min}$: maximum rotation frequency and associated minimum period.}\label{tab:MRfmax}
\end{table}

\subsubsection{Minimum mass and rotation frequency}\label{sec:4.2.2}

\begin{figure}[!hbtp]
\centering
\includegraphics[width=0.8\hsize,clip]{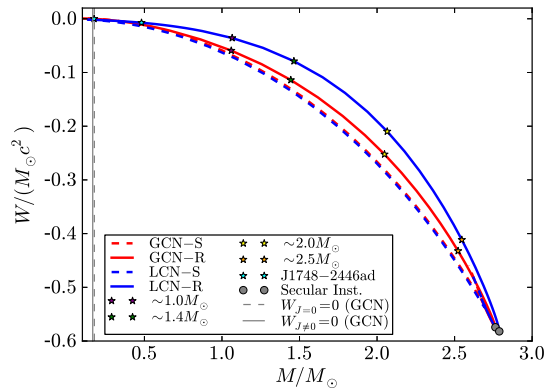}
\caption{Neutron star binding energy versus total mass along the Keplerian sequence both for the global (red) and local (blue) charge neutrality.}\label{fig:WM}
\end{figure}

We compute now the gravitational binding energy of the neutron star from Eq.~(\ref{eq:Wrot}) as a function of the central density and angular velocity. We make this for central densities higher than the nuclear density, thus we impose the neutron star to have a supranuclear hadronic core. In Fig.~\ref{fig:WM} we plot the binding energy $W$ of the neutron star as a function of the neutron star mass along the Keplerian sequence. For the sake of comparison we show also the binding energy of the non-rotating configurations.

We found that the globally neutral neutron stars studied here are bound up to some minimum mass at which the gravitational binding energy vanishes. For the static and Keplerian configurations we find that $W_{J=0}=0$, and $W_{J\neq 0}=0$ respectively at
\begin{equation}\label{eq:Mmin}
M^{J=0}_{\rm min}\approx 0.177\,M_\odot\;,\qquad M^{K}_{\rm min}\approx 0.167\,M_\odot,
\end{equation}
where with the superscript $K$ we indicate that this value corresponds to the minimum mass on the Keplerian sequence. Clearly this minimum mass value decreases with decreasing frequency until it reaches the above value $M^{J=0}_{\rm min}$ of the non-rotating case.

\begin{figure}[!hbtp]
\centering
\includegraphics[width=0.8\hsize,clip]{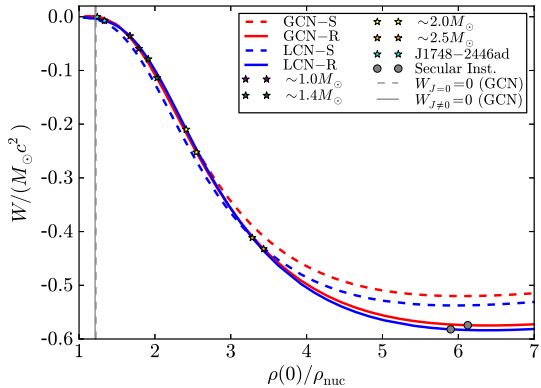}
\caption{Neutron star binding energy versus central density along the Keplerian sequence both for the global (red) and local (blue) charge neutrality.}\label{fig:Wrho}
\end{figure}

We did not find any unbound configuration in the local charge neutrality case for the present EOS (see Fig.~\ref{fig:WM}). The corresponding plot of $W$ as a function of the central density is shown in Fig.~\ref{fig:Wrho}.

The configuration with the minimum mass, $M^{K}_{\rm min}\approx 0.167\,M_\odot$, has a rotation frequency
\begin{equation}\label{eq:fmin}
f^{K}_{\rm min} = f (M^{K}_{\rm min})\approx 700.59\,{\rm Hz}\;,
\end{equation}
that is the minimum rotation rate that globally neutral configurations can have along the Keplerian sequence in order to be gravitationally bound. Interestingly, the above value is slightly lower than the frequency of the fastest observed pulsar, PSR J1748--2446ad, which has a frequency of 716 Hz \cite{hessels06}. Further discussions on this issue are given below in section~\ref{sec:8}.

\begin{figure}[!hbtp]
\centering
\includegraphics[width=0.8\hsize,clip]{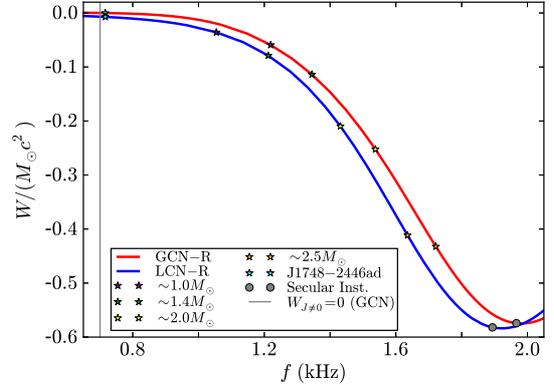}
\caption{Neutron star binding energy versus frequency for the Keplerian sequence both for the global (red) and local (blue) charge neutrality neutron stars.}\label{fig:Wf}
\end{figure}

In Fig.~\ref{fig:Wf} we show in detail the dependence of $W$ on the rotation frequency. 

\section{Neutron star mass-radius relation}\label{sec:5}

\begin{figure}[!hbtp]
\centering
\includegraphics[width=0.8\hsize]{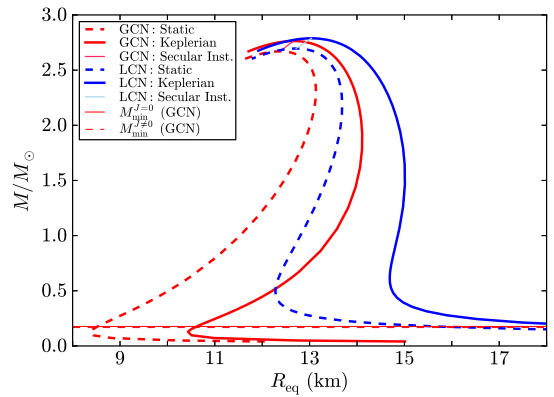}
\caption{Total mass versus total equatorial radius for the global (red) and local (blue) charge neutrality cases. The dashed curves represent the static configurations, while the solid lines are the uniformly rotating neutron stars. The pink-red and light-blue color lines define the secular instability boundary for the globally and locally neutral cases, namely the lines given by Eqs.~(\ref{eq:SecularG}) and (\ref{eq:SecularL}), respectively.}\label{fig:MtotvsRtot}
\end{figure}

We summarize now the above results in form of a new mass-radius relation of uniformly rotating neutron stars, including the Keplerian and secular instability boundary limits. In Fig.~\ref{fig:MtotvsRtot} we show a summary plot of the equilibrium configurations of rotating neutron stars. In particular we show the total mass versus the equatorial radius: the dashed lines represent the static (non-rotating, $J=0$) sequences, while the solid lines represent the corresponding Keplerian mass-shedding sequences. The secular instability boundaries are plotted in pink-red and light blue color for the global and local charge neutrality cases, respectively. 

It can be seen that due to the deformation for a given mass the radius of the rotating case is larger than the static one, and similarly the mass of the rotating star is larger than the corresponding static one. It can be also seen that the configurations obeying global charge neutrality are more compact with respect to the ones satisfying local charge neutrality. 

\section{Moment of inertia}\label{sec:6}

The neutron star moment of inertia $I$ can be computed from the relation
\begin{equation}\label{eq:I}
I = \frac{J}{\Omega}\;,
\end{equation}
where $J$ is the angular momentum and $\Omega$ are related via Eq.~(\ref{eq:Jomega}). Since $J$ is a first-order quantity and so proportional to $\Omega$, the moment of inertia given by Eq.~(\ref{eq:I}) does not depend on the angular velocity and does not take into account deviations from the spherical symmetry. This implies that Eq.~(\ref{eq:J}) gives the moment of inertia of the non-rotating unperturbed seed object. In order to find the perturbation to $I$, say $\delta I$, the perturbative treatment has to be extended to the next order $\Omega^3$, in such a way that $I=I_0+\delta I =(J_0+\delta J)/\Omega$, becomes of order $\Omega^2$, with $\delta J$ of order $\Omega^3$ \citep[see e.g.][]{1973Ap&SS..24..385H,benhar05}. In this work we keep the solution up to second order and therefore we proceed to analyze the behavior of the moment of inertia for the non-rotating configurations. In any case, as we will show in section~\ref{sec:8} even the fastest observed pulsars rotate at frequencies much lower than the Keplerian rate, and under such conditions we expect that the moment of inertia can be approximated with high accuracy by the one of the corresponding static configurations.

\begin{figure}[!hbtp]
\centering
\includegraphics[width=0.8\hsize,clip]{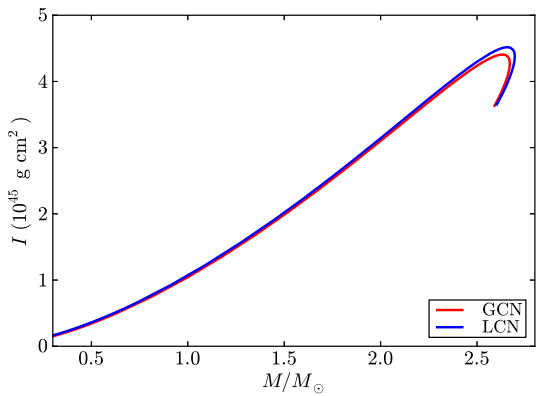}
\caption{Total moment of inertia versus total mass both for globally (red) and locally (blue) neutral non-rotating neutron stars.}\label{fig:inertiaM}
\end{figure}

\begin{figure}[!hbtp]
\centering
\includegraphics[width=0.8\hsize,clip]{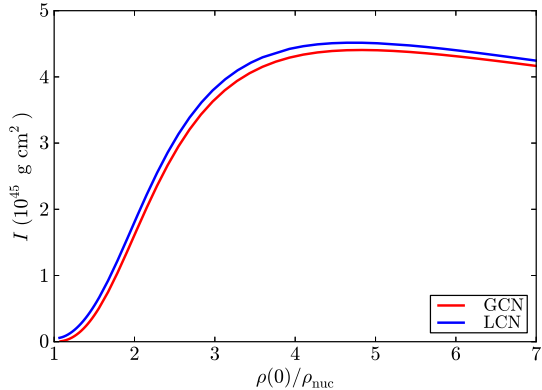}
\caption{Total moment of inertia versus central density for globally (red) and locally (blue) neutral non-rotating neutron stars.}\label{fig:inertiarho}
\end{figure}

In Figs.~\ref{fig:inertiaM} and \ref{fig:inertiarho} we show the behavior of the total momentum of inertia, i.e. $I=I_{\rm core}+I_{\rm crust}$, with respect to the total mass and central density for both globally and locally neutral non-rotating neutron stars.

We can see from Figs.~\ref{fig:inertiaM} and \ref{fig:inertiarho} that the total moment of inertia is quite similar for both global and local charge neutrality cases. This is due to the fact that the globally neutral configurations differ from the locally ones mostly in the structure of the crust, which however contributes much less than the neutron star core to the total moment of inertia (see below in section \ref{sec:4.4.1}).

\subsection{Core and crust moment of inertia}\label{sec:4.4.1}

In order to study the single contribution of the core and the crust to the moment of inertia of the neutron star, we shall use the integral expression for the moment of inertia. Multiplying Eq.~(\ref{eq:baromega}) by $r^3$ and making the integral of it we obtain\footnote{It is clear that this expression approaches, in the weak field limit, the classic Newtonian expression $I_{\rm Newtonian}=(8\pi/3)\int r^4 \rho\,dr$ where $\rho$ is the mass-density \citep{1967ApJ...150.1005H}.}
\begin{align}
I(r)&=-\frac{2}{3}\int_0^r r^3 \frac{d j}{dr}\frac{\bar{\omega}(r)}{\Omega}dr\nonumber \\
&=\frac{8\pi}{3}\int_0^r r^4 ({\cal E}+P) e^{(\lambda-\nu)/2}\frac{\bar{\omega}(r)}{\Omega}dr\;,
\end{align}
where the integration is carried out in the region of interest. Thus, the contribution of the core, $I_{\rm core}$, is obtained integrating from the origin up to the radius of the core, and the contribution of the crust, $I_{\rm crust}$, integrating from the base of the crust to the total radius of the neutron star. 

\begin{figure}[!hbtp]
\centering
\includegraphics[width=0.8\hsize,clip]{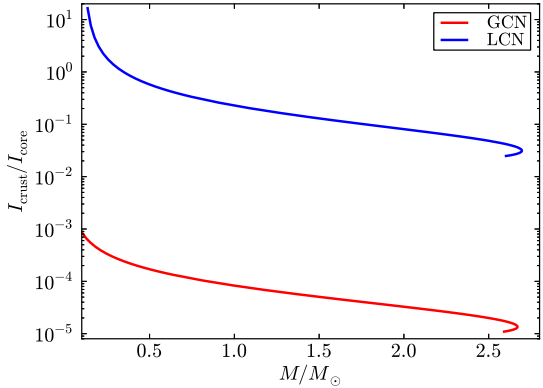}
\caption{Crust to core moment of inertia ratio versus the total mass of both globally and locally neutral non-rotating neutron stars.}\label{fig:incoreincrustM}
\end{figure}

\begin{figure}[!hbtp]
\centering
\includegraphics[width=0.8\hsize,clip]{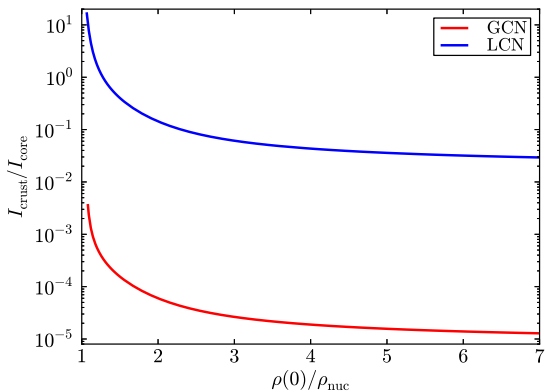}
\caption{Crust to core moment of inertia ratio versus the central density both globally and locally neutral non-rotating neutron stars.}\label{fig:incoreincrustrho}
\end{figure}

We show in Figs.~\ref{fig:incoreincrustM} and \ref{fig:incoreincrustrho} the ratio between the moment of inertia of the crust and the one of the core as a function of the total mass and central density, respectively, for both the globally and locally neutral configurations. 

\section{Deformation of the neutron star}\label{sec:7}

In this section we explore the deformation properties of the neutron star. The behavior of the eccentricity, the rotational to gravitational energy ratio, as well as the quadrupole moment, are investigated as a function of the mass, density, and rotation frequency of the neutron star.

\subsection{Eccentricity}\label{sec:4.5.1}

A measurement of the level of deformation of the neutron star can be estimated with the eccentricity
\begin{equation}\label{eq:eccentricity}
\epsilon=\sqrt{1-\left(\frac{R_p}{R_{\rm eq}}\right)^2}\;,
\end{equation}
where $R_p$ and $R_{\rm eq}$ are the polar and equatorial radii of the configuration. Thus, $\epsilon=0$ defines the spherical limit and $0<\epsilon<1$ corresponds to oblate configurations.

\begin{figure}[!hbtp]
\centering
\includegraphics[width=0.8\hsize,clip]{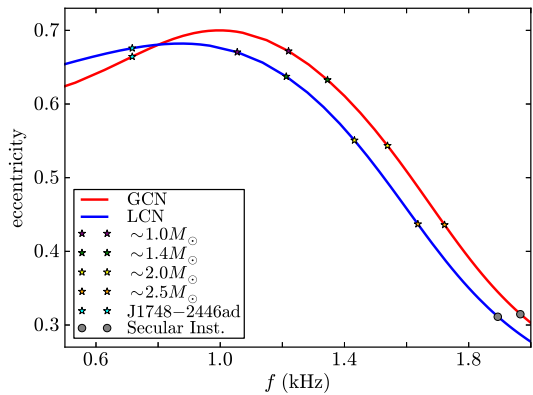}
\caption{Eccentricity (\ref{eq:eccentricity}) versus frequency for the Keplerian sequence both for the global (red) and local (blue) charge neutrality cases.}\label{fig:eccvsf}
\end{figure}

In Fig.~\ref{fig:eccvsf}, we show the behavior of the total eccentricity (\ref{eq:eccentricity}), as a function of the neutron star frequency.

\begin{figure}[!hbtp]
\centering
\includegraphics[width=0.8\hsize,clip]{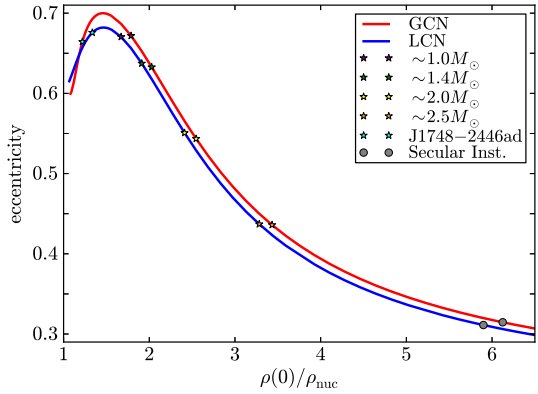}
\caption{Eccentricity (\ref{eq:eccentricity}) versus central density for the Keplerian sequence both for the global (red) and local (blue) charge neutrality cases.}\label{fig:eccvsrho}
\end{figure}

We can see that in general the globally neutral neutron star has an eccentricity larger than the one of the locally neutral configuration for almost the entire range of frequencies and the corresponding central densities, except for the low frequencies $f\lesssim 0.8$ kHz and central densities $\rho(0)\lesssim 1.3 \rho_{\rm nuc}$; see also Fig.~\ref{fig:eccvsrho}. Starting from low values of the frequency $f$ and central density $\rho(0)$, the neutron stars increase their oblateness, and after reaching the maximum value of the eccentricity, the compactness increases and the configurations tend to a more spherical shape.

\subsection{Rotational to gravitational energy ratio}\label{sec:4.5.2}

Other property of the star related to the centrifugal deformation of the star is the ratio between the gravitational energy and the rotational energy of the star. The former is given by Eq.~(\ref{eq:Wrot}), whereas the latter is
\begin{equation}
T = \frac{1}{2} I \Omega^2\;,
\end{equation}

\begin{figure}[!hbtp]
\centering
\includegraphics[width=0.8\hsize,clip]{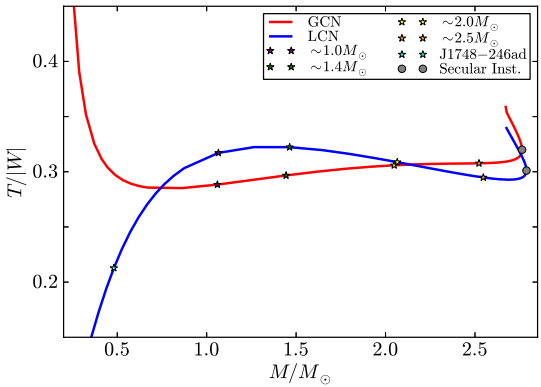}
\caption{Rotational to gravitational binding energy ratio versus total mass along the Keplerian sequence both for the global (red) and local (blue) charge neutrality.}\label{fig:ToverWM}
\end{figure}

\begin{figure}[!hbtp]
\centering
\includegraphics[width=0.8\hsize,clip]{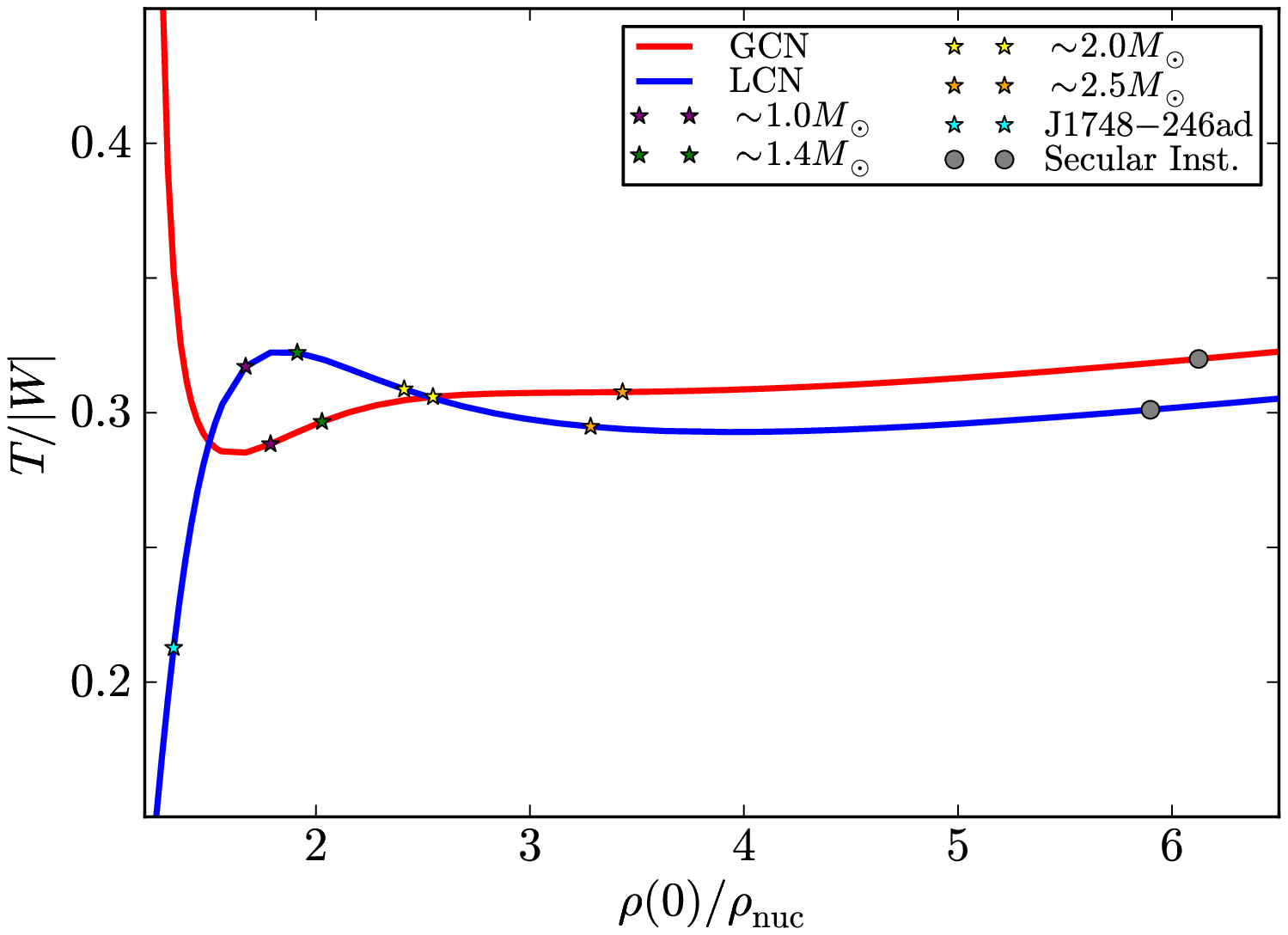}
\caption{Rotational to gravitational binding energy ratio versus central density along the Keplerian sequence both for the global (red) and local (blue) charge neutrality.}\label{fig:ToverWrho}
\end{figure}

\begin{figure}[!hbtp]
\centering
\includegraphics[width=0.8\hsize,clip]{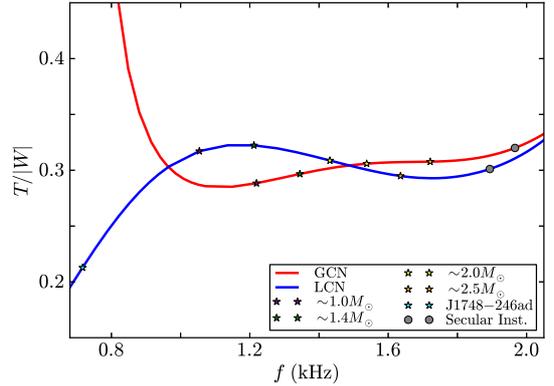}
\caption{Rotational to gravitational binding energy ratio versus frequency along the Keplerian sequence both for the global (red) and local (blue) charge neutrality cases.}\label{fig:ToverWf}
\end{figure}

We show in Fig.~\ref{fig:ToverWM} the ratio $T/|W|$ as a function of the mass of the neutron stars along the Keplerian sequence. In Fig.~\ref{fig:ToverWrho} instead we plot the dependence of the ratio on the central density and in Fig.~\ref{fig:ToverWf} on the Keplerian frequency.

\subsection{Quadrupole moment}\label{sec:4.4.3}

\begin{figure}[!hbtp]
\centering
\includegraphics[width=0.8\hsize,clip]{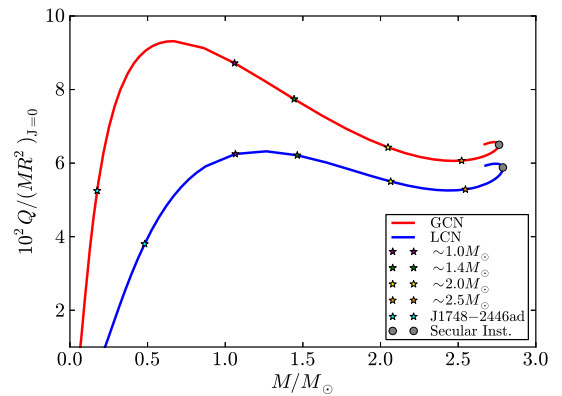}
\caption{Total quadrupole moment versus total mass along the Keplerian sequence both for the global (red) and local (blue) charge neutrality cases. The quadrupole moment $Q$ is here in units of the quantity $M R^2$ of the non-rotating configuration with the same central density.}\label{fig:QuadrupoleM}
\end{figure}

\begin{figure}[!hbtp]
\centering
\includegraphics[width=0.8\hsize,clip]{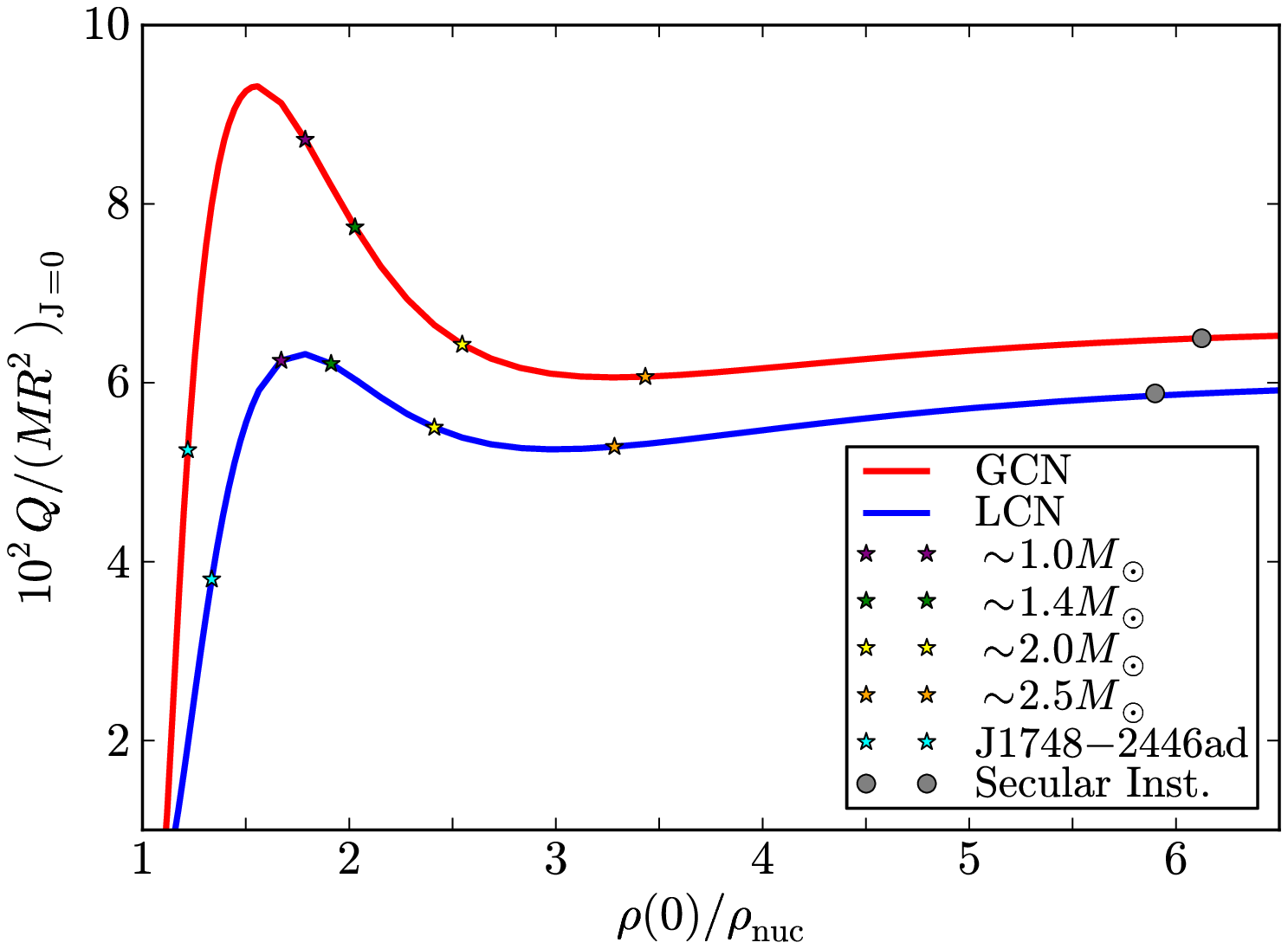}
\caption{Total quadrupole moment versus central density along the Keplerian sequence both for the global (red) and local (blue) charge neutrality cases. The quadrupole moment $Q$ is here in units of the quantity $M R^2$ of the non-rotating configuration with the same central density.}\label{fig:Quadrupolerho}
\end{figure}

\begin{figure}[!hbtp]
\centering
\includegraphics[width=0.8\hsize]{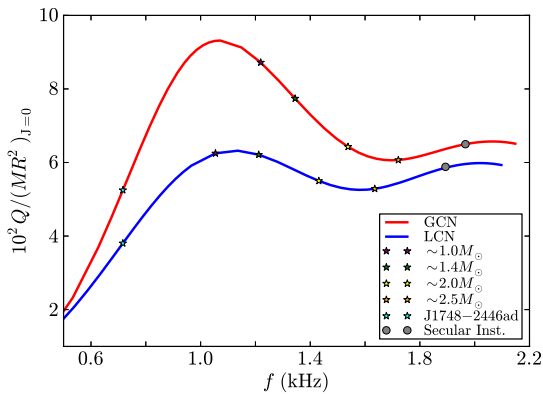}
\caption{Total quadrupole moment versus frequency along the Keplerian sequence both for the global (red) and local (blue) charge neutrality cases. The quadrupole moment $Q$ is here in units of the quantity $M R^2$ of the non-rotating configuration with the same central density.}\label{fig:Quadrupolef}
\end{figure}

In Figs.~\ref{fig:QuadrupoleM} and \ref{fig:Quadrupolerho} we show the quadrupole moment, $Q$ given by Eq.~(\ref{eq:Q}), as a function of the total mass and central density for both globally and locally neutral neutron stars along the Keplerian sequence. The dependence of $Q$ on the rotation frequency is shown in Fig.~\ref{fig:Quadrupolef}. We have normalized the quadrupole moment $Q$ to the quantity $M R^2$ of the non-rotating configuration with the same central density.

%

\section{Observational constraints}\label{sec:8}

\begin{figure}[!hbtp]
\centering
\includegraphics[width=0.8\hsize]{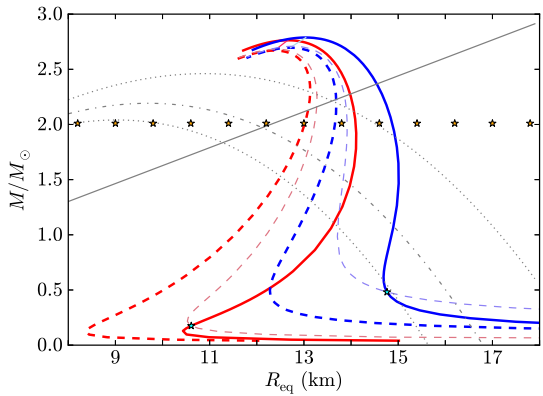}
\caption{Observational constraints on the mass-radius relation given by \cite{trumper2011} and the theoretical mass-radius relation presented in this work in Fig.~\ref{fig:MtotvsRtot}. The red lines represent the configuration with global charge neutrality, while the blue lines represent the configuration with local charge neutrality. The pink-red line and the light-blue line represent the secular axisymmetric stability boundaries for the globally neutral and the locally neutral case, respectively. The red and blue solid lines represent the Keplerian sequences and the red and blue dashed lines represent the static cases presented in \citep{2012NuPhA.883....1B}.}\label{fig:constraints}
\end{figure}

In Fig.~\ref{fig:constraints} we show the above mass-radius relations together with the most recent and stringent constraints indicated by \cite{trumper2011}: 

1) \emph{The largest mass}. Until 2013 it was given by the mass of the 3.15 millisecond pulsar PSR J1614-2230 $M=1.97 \pm 0.04 M_\odot$ \cite{demorest2010}, however the recent reported mass $2.01\pm 0.04 M_\odot$ for the neutron star in the relativistic binary PSR J0348+0432 \citep{2013Sci...340..448A} puts an even more stringent request to the nuclear EOS. Thus, the maximum mass of the neutron star has to be larger than the mass of PSR J0348+0432, this constraint is represented by the orange-color stars in Fig.~\ref{fig:constraints}.

2) \emph{The largest radius}. It is given by the lower limit to the radius of RX J1856-3754. The lower limit to the radius as seen by an observer at infinity is $R_\infty = R [1-2GM/(c^2 R)]^{-1/2} > 16.8$ km, as given by the fit of the optical and X-ray spectra of the source \cite{trumper04}; so in the mass-radius relation this constraint reads $2G M/c^2 >R-R^3/(R^{\rm min}_\infty)^2$, with $R^{\rm min}_\infty=16.8$ km. We represent this constraint with the dotted-dashed curve in Fig.~\ref{fig:constraints}.

3) \emph{The maximum surface gravity}. Using a neutron star of $M=1.4M_\odot$ to fit the Chandra data of the low-mass X-ray binary X7, it turns out that the radius of the star satisfies at 90$\%$ confidence level, $R=14.5^{+1.8}_{-1.6}$ km, which gives $R_\infty = [15.64,18.86]$ km, respectively \cite{heinke06}. Using the same formula as before, $2G M/c^2 >R-R^3/(R^{\rm min}_\infty)^2$, we obtain the dotted curves shown in Fig.~\ref{fig:constraints}.

4) \emph{The highest rotation frequency}. The fastest observed pulsar is PSR J1748--2446ad with a frequency of 716 Hz \cite{hessels06}. We show the constant rotation frequency sequence $f=716$ Hz for both globally (dashed pink) and locally (dashed light blue) neutral neutron stars. We indicated with cyan-color stars the point where these curves cross the corresponding Keplerian sequences in the two cases (see Fig.~\ref{fig:constraints}).

Every $f$-constant sequence crosses the stability region of the objects in two points: these crossing points define the minimum and maximum possible mass that an object rotating with such a frequency may have in order to be stable. In the case of PSR J1748-2446ad, the cut of the $f=716$ Hz constant sequence with the Keplerian curve establishes the minimum mass of this pulsar. We find that its minimum mass is $\approx 0.175\,M_\odot$ and corresponding equatorial radius $10.61$ km for the globally neutral neutron star. For the locally neutral configuration we found $\approx 0.48\,M_\odot$ and $14.8$ km, respectively for the minimum mass and corresponding equatorial radius. This implies that the mass of PSR J1748-2446ad is poorly constrained to be larger than the above values.

It is interesting that the above minimum mass, given by its constant rotation frequency sequence, is slightly larger than the minimum mass for bound configurations on the Keplerian sequence, $M^{K}_{\rm min}\approx 0.167\,M_\odot$; see Eq.~(\ref{eq:Mmin}). In fact, as we shown in Eq.~(\ref{eq:fmin}) the minimum rotation frequency along the Keplerian sequence for bound configurations in the globally neutral case is, $f^{K}_{\rm min}\approx 700.59$ Hz, which is slightly lower than the frequency of PSR J1748-2446ad. It would imply that PSR J1748-2446ad is very likely rotating at a rate much lower than the Keplerian one.

Similarly to what presented in \cite{2012NuPhA.883....1B} for the static neutron stars and introduced by \cite{trumper2011}, the above observational constraints show a preference on stiff EOS that provide highest maximum masses for neutron stars. Taking into account the above constraints, the radius of a canonical neutron star of mass $M=1.4 M_\odot$ is strongly constrained to $R\geq12$ km, disfavoring at the same time strange quark matter stars. It is evident from Fig.~\ref{fig:constraints} that mass-radius relations for both the static and the rotating case presented here, are consistent with all the observational constraints. In Table \ref{tab:MRprediction} we show the radii predicted by our mass-radius relation both for the static and the rotating case for a canonical neutron star as well as for the most massive neutron stars discovered, namely, the millisecond pulsar PSR J1614--2230 \cite{demorest2010}, $M=1.97 \pm 0.04 M_\odot$, and the most recent PSR J0348+0432, $M=2.01 \pm 0.04 M_\odot$ \citep{2013Sci...340..448A}.

\begin{table}
\centering
\begin{tabular}{c c c}
$M (M_\odot)$ & $R^{J=0}$ (km)	& $R^{J\neq0}_{\rm eq}$ (km)\\
\hline
1.40 & 12.313 & 13.943 \\

1.97 & 12.991 & 14.104 \\

2.01 & 13.020 & 14.097 \\

\hline
\end{tabular}
\caption{Radii for a canonical neutron star of $M=1.4 M_\odot$ and for PSR J1614--2230 \cite{demorest2010}, $M=1.97 \pm 0.04 M_\odot$, and PSR J0348+0432 \citep{2013Sci...340..448A}, $M=2.01 \pm 0.04 M_\odot$. These configurations are computed under the constraint of global charge neutrality and for a density at the edge of the crust equal to the neutron drip density. The nuclear parameterizations NL3 has been used.}\label{tab:MRprediction}
\end{table}

%

\section{Concluding remarks}\label{sec:9}

We have constructed equilibrium configurations of uniformly rotating neutron stars in both the global charge neutrality and local charge neutrality cases, generalizing our previous work \citep{2012NuPhA.883....1B}. To do this we have applied the Hartle method to the seed static solution obtained from the integration of the Einstein-Maxwell-Thomas-Fermi equations \citep{2012NuPhA.883....1B}. We calculated the mass, polar and equatorial radii, angular momentum, moment of inertia, quadrupole moment, and eccentricity, as functions of the central density and the rotation angular velocity of the neutron star.

The Keplerian mass-shedding limit and the secular axisymmetric instability have been analyzed for the construction of the region of stability of rotating neutron stars. We have given fitting curves of the secular instability boundary in Eqs.~(\ref{eq:SecularG}) and (\ref{eq:SecularL}) for global and local charge neutrality, respectively. With this analysis we have established in section~\ref{sec:4.2.1} the maximum mass and maximum rotation frequency of the neutron star. We computed in section~\ref{sec:4.2.2} the gravitational binding energy of the configurations as a function of the central density and rotation rate. We did this for central densities higher than the nuclear one, so imposing that the neutron star has a supranuclear hadronic core. We found that there is a minimum mass under which the neutron star becomes gravitationally unbound. Along the Keplerian sequence, to this minimum mass object we associate a minimum frequency under which an object rotating at the Keplerian rate becomes unbound; see Eq.~(\ref{eq:fmin}). We found that locally neutral neutron stars with supranuclear cores remained always bound for the present EOS. In Table~\ref{tab:conclusions} we summarize all these results.

\begin{table}
\centering
\begin{tabular}{c c c}
& Global Neutrality & Local Neutrality \\
\hline
$M^{J=0}_{\rm max} (M_\odot)$ & 2.67 & 2.70\\
$M^{J\neq 0}_{\rm max} (M_\odot)$ & 2.76 & 2.79\\
$f_{\rm max}$ (kHz)& 1.97 & 1.89\\
$P_{\rm min}$ (ms)& 0.51 & 0.53\\
$M^{J= 0}_{\rm min} (M_\odot)$ & 0.18 & --\\
$M^{K}_{\rm min} (M_\odot)$ & 0.17 & --\\
$f^{K}_{\rm min}$ (kHz) & 0.70 & --\\
\hline
\end{tabular}
\caption{Maximum mass, maximum frequency, minimum period, minimum mass of globally and locally neutral neutron stars.}\label{tab:conclusions}
\end{table}

We finally analyzed in section~\ref{sec:8} the current observational constraints on the mass-radius relation of neutron stars. We constructed the constant frequency sequence of PSR J1748--2446ad to obtain the minimum possible mass of this source, which is given by the crossing point of the $f=716$ Hz constant sequence with the Keplerian one. It gives $\approx 0.17\,M_\odot$ and $\approx 0.48\,M_\odot$ for the global and charge neutrality cases, respectively. The very low mass inferred for PSR J1748--2446ad assuming that it rotates at the Keplerian rate implies that its frequency is unlikely to be actually the Keplerian. Otherwise, it would imply that PSR J1748--32446ad could be the less massive neutron star ever observed.

It would be interesting to analyze the generality of the neutron star features shown in this work since the most recent measurement of the mass PSR J0348+0432, $M=2.01 \pm 0.04 M_\odot$ \citep{2013Sci...340..448A}, favors stiff nuclear EOS as the one used here.

\section*{References}



\appendix

\section{The Hartle solution and equatorial circular orbits}\label{app:1}

\subsection{The Hartle-Thorne vacuum solution}\label{app:1a}

It is possible to write the Hartle-Thorne metric given by eq.~\ref{eq:rotmetric} in an analytic closed-form in the exterior vacuum case as function of the total mass $M$, angular momentum $J$, and quadrupole moment $Q$ of the rotating star.The angular velocity of local inertial frames $\omega(r)$, proportional to $\Omega$, and the functions $h_0$, $h_2$, $m_0$, $m_2$, $k_2$, proportional to $\Omega^2$, are derived from the Einstein equations \citep[see][for details]{1967ApJ...150.1005H,1968ApJ...153..807H}.  Following this prescriptions the eq.~\ref{eq:rotmetric} become:
\begin{align}\label{ht1}
&ds^2=\left(1-\frac{2{ M }}{r}\right)\bigg[1+2k_1P_2(\cos\theta)\nonumber\\
&+2\left(1-\frac{2{ M}}{r}\right)^{-1}\frac{J^{2}}{r^{4}}(2\cos^2\theta-1)\bigg]dt^2\nonumber\\
&+\frac{4J}{r}\sin^2\theta dt d\phi-\left(1-\frac{2{ M}}{r}\right)^{-1}\nonumber\\
&\times\bigg[1-2\left(k_1-\frac{6 J^{2}}{r^4}\right)P_2(\cos\theta) \nonumber \\
&-2\left(1-\frac{2{ M}}{r}\right)^{-1}\frac{J^{2}}{r^4}\bigg]dr^2\nonumber\\
&-r^2[1-2k_2P_2(\cos\theta)](d\theta^2+\sin^2\theta d\phi^2),
\end{align}

where
\begin{align*}
k_1&=\frac{J^2}{M r^3}\left(1+\frac{M}{r}\right)+\frac{5}{8}\frac{Q-J^{2}/{M}}{M^3}Q_2^2(x) \;,\\
k_2&=k_1+\frac{J^{2}}{r^4}+\frac{5}{4}\frac{Q-J^{2}/{ M}}{{ M}^2r \sqrt{1-2M/r}}Q_2^1(x) \;,
\end{align*}
and
\begin{align*}\label{legfunc}
Q_2^1(x)&=(x^2-1)^{1/2}\left[\frac{3x}{2}\ln\left(\frac{x+1}{x-1}\right)-\frac{3x^2-2}{x^2-1}\right] \;, \\
Q_2^2(x)&=(x^2-1)\left[\frac{3}{2}\ln\left(\frac{x+1}{x-1}\right)-\frac{3x^3-5x}{(x^2-1)^2}\right] \;,
\end{align*}
are the associated Legendre functions of the second kind, being $P_2(\cos\theta)=(1/2)(3\cos^2\theta-1)$ the Legendre polynomial, and where it has been effectuated the re-scaling $x=r/M -1$. The constants $M$, $J$ and $Q$ are the total mass, angular momentum and mass quadrupole moment of the rotating object, respectively. This form of the metric corrects some misprints of the original paper by \cite{1968ApJ...153..807H} \citep[see also][]{2005MNRAS.358..923B,2012PhRvD..86f4043B}. To obtain the exact numerical values of $M$, $J$ and $Q$, the exterior and interior metrics have to be matched at the surface of the star. It is worthy underline that in the terms involving $J^2$ and $Q$, the total mass $M$ can be substituted by $M^{J=0}$ since $\delta M$ is already a second order term in the angular velocity.

\subsection{Angular velocity of equatorial circular orbits}\label{app:1b}

It is possible to obtain the analytical expression for the angular velocity $\Omega$ given by Eq.~(\ref{eq:omegaKep}) with respect to an observer at infinity, taking into account the parameterization of the four-velocity $u$ of a test particle on a circular orbit  in equatorial plane of axisymmetric stationary spacetime, regarding as parameter the angular velocity $\Omega$ itself:
\begin{equation}
u=\Gamma[\partial_t+\Omega\partial_{\phi}] \;,
\end{equation}
where $\Gamma$ is a normalization factor such that $u^{\alpha}u_{\alpha}=1$. Normalizing and applying the geodesics conditions we get the following expressions for $\Gamma$ and $\Omega=u^{\phi}/u^{t}$
\begin{align}
&\Gamma=\pm(g_{tt}+2\Omega g_{t\phi}+\Omega^2 g_{\phi\phi})^{-1/2} \;,\label{eight}\\
&g_{tt,r}+2\Omega g_{t\phi,r}+\Omega^2 g_{\phi\phi,r}=0 \label{eight1}\;.
\end{align}
Thus, the solution of Eqs.~(\ref{eight}--\ref{eight1}) can be written as
\begin{equation}
\Omega^\pm_{\rm orb}(r)=\frac{u^{\phi}}{u^{t}}=\frac{-g_{t\phi,r}\pm\sqrt{(g_{t\phi,r})^2-g_{tt,r}g_{\phi\phi,r}}}{g_{\phi\phi,r}} \;,
\end{equation}
where $+/-$ stands for co-rotating/counter-rotating orbits, $u^{\phi}$ and $u^{t}$ are the angular and time components of the four-velocity respectively, and a colon stands for partial derivative with respect to the corresponding coordinate. To determine the mass shedding angular velocity (the Keplerian angular velocity) of the neutron stars, we need to consider only the co-rotating orbit, so from here and thereafter we take into account only the plus sign in Eq.~(\ref{eight}) and we write $\Omega^{+}_{\rm orb}(r)=\Omega_{\rm orb}(r)$.

For the Hartle external solution given by Eq.~(\ref{ht1}) we obtain Eq.~(\ref{eq:omegaKep}) with
\begin{align*}
F_1&=\left(\frac{M}{r}\right)^{3/2} \;,\\
F_2&=\frac{48M^7-80M^6r+4M^5r^2-18M^4r^3}{16M^2r^4(r-2M)}\\
&+\frac{40M^3r^4+10M^2r^5+15Mr^6-15r^7}{16M^2r^4(r-2M)}+F\;,\\
F_3&=\frac{6M^4-8M^3r-2M^2r^2-3Mr^3+3r^4}{16M^2r(r-2M)/5}-F \;,\\
F&=\frac{15(r^3-2M^3)}{32M^3}\ln\frac{r}{r-2M} \;.\\
\end{align*}

The maximum angular velocity possible for a rotating star at the mass-shedding limit is the Keplerian angular velocity evaluated at the equator ($r=R_{\rm eq}$), i.e.
\begin{equation}\label{eq:omegaK1}
\Omega_K^{J\neq0}=\Omega_{\rm orb}(r=R_{\rm eq})\;.
\end{equation}
In the static case i.e. when $j=0$ hence $q=0$ and $\delta M=0$ we have the well-known Schwarzschild solution and the orbital angular velocity for a test particle $\Omega_K^{J=0}$ on the surface ($r=R$) of the neutron star is given by
\begin{equation}\label{eq:omegaK2}
\Omega_K^{J=0}=\sqrt{\frac{M^{J=0}}{R_{M^{J=0}}^3}}\;.
\end{equation}

\begin{thebibliography}{34}
\expandafter\ifx\csname natexlab\endcsname\relax\def\natexlab#1{#1}\fi
\expandafter\ifx\csname url\endcsname\relax
  \def\url#1{\texttt{#1}}\fi
\expandafter\ifx\csname urlprefix\endcsname\relax\def\urlprefix{URL }\fi

\bibitem[{{Antoniadis} et~al.(2013){Antoniadis}, {Freire}, {Wex}, {Tauris},
  {Lynch}, {van Kerkwijk}, {Kramer}, {Bassa}, {Dhillon}, {Driebe}, {Hessels},
  {Kaspi}, {Kondratiev}, {Langer}, {Marsh}, {McLaughlin}, {Pennucci}, {Ransom},
  {Stairs}, {van Leeuwen}, {Verbiest}, and {Whelan}}]{2013Sci...340..448A}
{Antoniadis}, J., {Freire}, P.~C.~C., {Wex}, N., {Tauris}, T.~M., {Lynch},
  R.~S., {van Kerkwijk}, M.~H., {Kramer}, M., {Bassa}, C., {Dhillon}, V.~S.,
  {Driebe}, T., {Hessels}, J.~W.~T., {Kaspi}, V.~M., {Kondratiev}, V.~I.,
  {Langer}, N., {Marsh}, T.~R., {McLaughlin}, M.~A., {Pennucci}, T.~T.,
  {Ransom}, S.~M., {Stairs}, I.~H., {van Leeuwen}, J., {Verbiest}, J.~P.~W.,
  {Whelan}, D.~G., Apr. 2013. {A Massive Pulsar in a Compact Relativistic
  Binary}. Science 340, 448.

\bibitem[{{Belvedere} et~al.(2012){Belvedere}, {Pugliese}, {Rueda}, {Ruffini},
  and {Xue}}]{2012NuPhA.883....1B}
{Belvedere}, R., {Pugliese}, D., {Rueda}, J.~A., {Ruffini}, R., {Xue}, S.-S.,
  Jun. 2012. {Neutron star equilibrium configurations within a fully
  relativistic theory with strong, weak, electromagnetic, and gravitational
  interactions}. Nuclear Physics A 883, 1--24.

\bibitem[{{Benhar} et~al.(2005){Benhar}, {Ferrari}, {Gualtieri}, and
  {Marassi}}]{benhar05}
{Benhar}, O., {Ferrari}, V., {Gualtieri}, L., {Marassi}, S., Aug. 2005.
  {Perturbative approach to the structure of rapidly rotating neutron stars}.
  \prd 72~(4), 044028--+.

\bibitem[{{Berti} et~al.(2005){Berti}, {White}, {Maniopoulou}, and
  {Bruni}}]{2005MNRAS.358..923B}
{Berti}, E., {White}, F., {Maniopoulou}, A., {Bruni}, M., Apr. 2005. {Rotating
  neutron stars: an invariant comparison of approximate and numerical
  space-time models}. \mnras 358, 923--938.

\bibitem[{{Bini} et~al.(2013){Bini}, {Boshkayev}, {Ruffini}, and
  {Siutsou}}]{BBRS2013}
{Bini}, D., {Boshkayev}, K., {Ruffini}, R., {Siutsou}, I., 2013. {Equatorial
  circular geodesics in the Hartle-Thorne spacetime}. Il Nuovo Cimento C 36,
  31.

\bibitem[{{Boguta} and {Bodmer}(1977)}]{boguta77}
{Boguta}, J., {Bodmer}, A.~R., Dec. 1977. {Relativistic calculation of nuclear
  matter and the nuclear surface}. Nuclear Physics A 292, 413--428.

\bibitem[{{Boshkayev} et~al.(2012{\natexlab{a}}){Boshkayev}, {Quevedo}, and
  {Ruffini}}]{2012PhRvD..86f4043B}
{Boshkayev}, K., {Quevedo}, H., {Ruffini}, R., Sep. 2012{\natexlab{a}}.
  {Gravitational field of compact objects in general relativity}. \prd 86~(6),
  064043.

\bibitem[{{Boshkayev} et~al.(2012{\natexlab{b}}){Boshkayev}, {Rotondo}, and
  {Ruffini}}]{2012IJMPS..12...58B}
{Boshkayev}, K., {Rotondo}, M., {Ruffini}, R., Mar. 2012{\natexlab{b}}. {On
  Magnetic Fields in Rotating Nuclear Matter Cores of Stellar Dimensions}.
  International Journal of Modern Physics Conference Series 12, 58--67.

\bibitem[{{Demorest} et~al.(2010){Demorest}, {Pennucci}, {Ransom}, {Roberts},
  and {Hessels}}]{demorest2010}
{Demorest}, P.~B., {Pennucci}, T., {Ransom}, S.~M., {Roberts}, M.~S.~E.,
  {Hessels}, J.~W.~T., Oct. 2010. {A two-solar-mass neutron star measured using
  Shapiro delay}. \nat 467, 1081--1083.

\bibitem[{{Friedman} et~al.(1988){Friedman}, {Ipser}, and
  {Sorkin}}]{1988ApJ...325..722F}
{Friedman}, J.~L., {Ipser}, J.~R., {Sorkin}, R.~D., Feb. 1988. {Turning-point
  method for axisymmetric stability of rotating relativistic stars}. \apj 325,
  722--724.

\bibitem[{{Friedman} et~al.(1986){Friedman}, {Parker}, and
  {Ipser}}]{Friedman1986}
{Friedman}, J.~L., {Parker}, L., {Ipser}, J.~R., May 1986. {Rapidly rotating
  neutron star models}. \apj 304, 115--139.

\bibitem[{{Haensel} et~al.(2007){Haensel}, {Potekhin}, and
  {Yakovlev}}]{haenselbook}
{Haensel}, P., {Potekhin}, A.~Y., {Yakovlev}, D.~G. (Eds.), 2007. {Neutron
  Stars 1 : Equation of State and Structure}. Vol. 326 of Astrophysics and
  Space Science Library.

\bibitem[{{Hartle}(1967)}]{1967ApJ...150.1005H}
{Hartle}, J.~B., Dec. 1967. {Slowly Rotating Relativistic Stars. I. Equations
  of Structure}. \apj 150, 1005.

\bibitem[{{Hartle}(1973)}]{1973Ap&SS..24..385H}
{Hartle}, J.~B., Oct. 1973. {Slowly Rotating Relativistic Stars. IX: Moments of
  Inertia of Rotationally Distorted Stars}. \apss 24, 385--405.

\bibitem[{{Hartle} and {Sharp}(1967)}]{HS1967}
{Hartle}, J.~B., {Sharp}, D.~H., Jan. 1967. {Variational Principle for the
  Equilibrium of a Relativistic, Rotating Star}. \apj 147, 317--+.

\bibitem[{{Hartle} and {Thorne}(1968)}]{1968ApJ...153..807H}
{Hartle}, J.~B., {Thorne}, K.~S., Sep. 1968. {Slowly Rotating Relativistic
  Stars. II. Models for Neutron Stars and Supermassive Stars}. \apj 153, 807.

\bibitem[{{Heinke} et~al.(2006){Heinke}, {Rybicki}, {Narayan}, and
  {Grindlay}}]{heinke06}
{Heinke}, C.~O., {Rybicki}, G.~B., {Narayan}, R., {Grindlay}, J.~E., Jun. 2006.
  {A Hydrogen Atmosphere Spectral Model Applied to the Neutron Star X7 in the
  Globular Cluster 47 Tucanae}. \apj 644, 1090--1103.

\bibitem[{{Hessels} et~al.(2006){Hessels}, {Ransom}, {Stairs}, {Freire},
  {Kaspi}, and {Camilo}}]{hessels06}
{Hessels}, J.~W.~T., {Ransom}, S.~M., {Stairs}, I.~H., {Freire}, P.~C.~C.,
  {Kaspi}, V.~M., {Camilo}, F., Mar. 2006. {A Radio Pulsar Spinning at 716 Hz}.
  Science 311, 1901--1904.

\bibitem[{{Klein}(1949)}]{klein49}
{Klein}, O., Jul. 1949. {On the Thermodynamical Equilibrium of Fluids in
  Gravitational Fields}. Reviews of Modern Physics 21, 531--533.

\bibitem[{{Lalazissis} et~al.(1997){Lalazissis}, {K{\"o}nig}, and
  {Ring}}]{lalazissis97}
{Lalazissis}, G.~A., {K{\"o}nig}, J., {Ring}, P., Jan. 1997. {New
  parametrization for the Lagrangian density of relativistic mean field
  theory}. \prc 55, 540--543.

\bibitem[{{Oppenheimer} and {Volkoff}(1939)}]{oppenheimer39}
{Oppenheimer}, J.~R., {Volkoff}, G.~M., Feb. 1939. {On Massive Neutron Cores}.
  \pr 55, 374--381.

\bibitem[{{Rotondo} et~al.(2011){Rotondo}, {Rueda}, {Ruffini}, and
  {Xue}}]{2011PhLB..701..667R}
{Rotondo}, M., {Rueda}, J.~A., {Ruffini}, R., {Xue}, S.-S., Jul. 2011. {The
  self-consistent general relativistic solution for a system of degenerate
  neutrons, protons and electrons in {$\beta$}-equilibrium}. Physics Letters B
  701, 667--671.

\bibitem[{{Rueda} et~al.(2011){Rueda}, {Ruffini}, and
  {Xue}}]{2011NuPhA.872..286R}
{Rueda}, J.~A., {Ruffini}, R., {Xue}, S.-S., Dec. 2011. {The Klein first
  integrals in an equilibrium system with electromagnetic, weak, strong and
  gravitational interactions}. Nuclear Physics A 872, 286--295.

\bibitem[{{Shapiro} and {Teukolsky}(1983)}]{shapirobook}
{Shapiro}, S.~L., {Teukolsky}, S.~A., 1983. {Black holes, white dwarfs, and
  neutron stars: The physics of compact objects}.

\bibitem[{{Sorkin}(1981)}]{1981ApJ...249..254S}
{Sorkin}, R., Oct. 1981. {A Criterion for the Onset of Instability at a Turning
  Point}. \apj 249, 254.

\bibitem[{{Sorkin}(1982)}]{1982ApJ...257..847S}
{Sorkin}, R.~D., Jun. 1982. {A Stability Criterion for Many Parameter
  Equilibrium Families}. \apj 257, 847.

\bibitem[{{Stergioulas}(2003)}]{2003LRR.....6....3S}
{Stergioulas}, N., Jun. 2003. {Rotating Stars in Relativity}. Living Reviews in
  Relativity 6, 3.

\bibitem[{{Takami} et~al.(2011){Takami}, {Rezzolla}, and
  {Yoshida}}]{2011MNRAS.416L...1T}
{Takami}, K., {Rezzolla}, L., {Yoshida}, S., Sep. 2011. {A quasi-radial
  stability criterion for rotating relativistic stars}. \mnras 416, L1--L5.

\bibitem[{{Tolman}(1930)}]{1930PhRv...35..904T}
{Tolman}, R.~C., Apr. 1930. {On the Weight of Heat and Thermal Equilibrium in
  General Relativity}. Physical Review 35, 904--924.

\bibitem[{{Tolman}(1939)}]{tolman39}
{Tolman}, R.~C., Feb. 1939. {Static Solutions of Einstein's Field Equations for
  Spheres of Fluid}. Physical Review 55, 364--373.

\bibitem[{{Torok} et~al.(2008){Torok}, {Bakala}, {Stuchlik}, and
  {Cech}}]{2008AcA....58....1T}
{Torok}, G., {Bakala}, P., {Stuchlik}, Z., {Cech}, P., Mar. 2008. {Modeling the
  Twin Peak QPO Distribution in the Atoll Source 4U 1636-53}. Acta Astronomica
  58, 1--14.

\bibitem[{{Tr{\"u}mper}(2011)}]{trumper2011}
{Tr{\"u}mper}, J.~E., Jul. 2011. {Observations of neutron stars and the
  equation of state of matter at high densities}. Progress in Particle and
  Nuclear Physics 66, 674--680.

\bibitem[{{Tr{\"u}mper} et~al.(2004){Tr{\"u}mper}, {Burwitz}, {Haberl}, and
  {Zavlin}}]{trumper04}
{Tr{\"u}mper}, J.~E., {Burwitz}, V., {Haberl}, F., {Zavlin}, V.~E., Jun. 2004.
  {The puzzles of RX J1856.5-3754: neutron star or quark star?} Nuclear Physics
  B Proceedings Supplements 132, 560--565.

\bibitem[{{Weber} and {Glendenning}(1992)}]{1992ApJ...390..541W}
{Weber}, F., {Glendenning}, N.~K., May 1992. {Application of the improved
  Hartle method for the construction of general relativistic rotating neutron
  star models}. \apj 390, 541--549.

\end{thebibliography}
\end{document}